\newif\ifPDF
\documentclass[usenatbib]{mn2e}
\ifPDF
  \usepackage[T1]{fontenc}
  \usepackage{times}
  \usepackage[hypertex,colorlinks=true,citecolor=blue]{hyperref}
  \usepackage{aeguill}
  \usepackage[pdftex]{graphicx,color}
  \usepackage{subfigure}
  \usepackage{afterpage}
\else
  \usepackage[T1]{fontenc}
  \usepackage{times}
  \usepackage[dvips]{graphicx,color}
  \usepackage{subfigure}
  \usepackage{afterpage}
\fi
\title[Investigation of the unique nulling properties of PSR B0818$-$41]
{Investigation of the unique nulling properties of PSR B0818$-$41}
\author[Bhaswati Bhattacharyya]
{Bhaswati Bhattacharyya$^1$,
 Yashwant Gupta$^2$, Janusz Gil$^3$\\\\
 $^1$Inter-University Centre for Astronomy and Astrophysics, Pune University Campus, Pune 411 007, India\\
 $^2$National Centre for Radio Astrophysics, TIFR, Pune University Campus, Post Bag 3,
Pune 411 007, India\\
 $^3$Institute of Astronomy, University of Zielona Gora, Lubuska 2, 65-265 Zielona Gora, Poland }

\date{Accepted. Received}
\begin{document}
\label{firstpage}
\maketitle
\pagerange{\pageref{firstpage}--\pageref{lastpage}} \pubyear{2009}
\def\LaTeX{L\kern-.36em\raise.3ex\hbox{a}\kern-.15em
    T\kern-.1667em\lower.7ex\hbox{E}\kern-.125emX}

\begin{abstract}
We report on the unique nulling properties of PSR B0818$-$41, using the 
GMRT at 325 and 610 MHz. This pulsar shows well defined nulls, with lengths
ranging from a few tens of pulses to a few hundreds of pulses.  We estimate
a nulling fraction of about 30\% at 325 MHz.  Furthermore, we find the 
following interesting behaviour of the pulse intensities, pulse shapes,
pulse widths and the drift rate, just before and after the nulls:
(i) There is a clear difference between the transitions from bursts to nulls 
to that from the nulls to bursts. The pulsar's intensity does not switch off 
abruptly at the null, but fades gradually, taking $\sim$ 10$P_1$. On the 
other hand, just after the nulls the intensity rises to a maximum over a 
short (less than one period) time scale. 
(ii) While the last active pulses before nulls are dimmer, the first few 
active pulses just after the nulls outshine the normal ones. This effect is 
very clear for the inner region of the pulsar profile, where the mean intensity
of the last few active pulses just after the nulls is $\sim$ 2.8 times more 
than that for the last active pulses just before the nulls.   
(iii) There is a significant evolution of the shape of the pulsar's profile, 
around the nulls, especially at the beginning of the bursts: an enhanced 
bump of intensity in the inner region, a change in the ratio of the strengths
of the leading and trailing peaks towards a more symmetric profile, an 
increase in profile width of about 10\%, and a shift of the profile centre
towards later longitudes.  Some of these can be explained by a (temporary?) 
shift of the emission regions to different heights and/or slightly outer 
field lines in the magnetosphere.
(iv) Just before the onset of the nulls, for about 60\% of the occasions, the 
apparent drift rate becomes slower (correlated with the gradual decrease of 
pulse intensity), transitioning to an almost phase stationary drift pattern. 
Further, when the pulsar comes out of the null, the increased intensity is very 
often accompanied by what looks like a disturbed drift rate behaviour, which 
settles down to the regular drift pattern as the pulsar intensity returns to 
normal.  Thus, we find some very specific and well correlated changes in the 
radio emission properties of PSR B0818$-$41 when the emission restarts after a 
null.  These could imply that the phenomenon of nulling is associated with some 
kind of a ``reset'' of the pulsar radio emission engine. 
We also present plausible explanations for some of the observed behaviour, using 
the Partially Screened Gap model of the inner pulsar accelerator.
\end{abstract}

\label{firstpage} \pagerange{\pageref{firstpage}--\pageref{lastpage}} %
\pubyear{2009}


\section{Introduction}

\label{ch4_intro} Many pulsars are known to exhibit the phenomenon of
nulling, where the emission appears to cease, or is greatly diminished, for
a certain number of pulse periods. Nulling is considered an important clue 
towards unraveling the mystery of the pulsar emission mechanism. Detailed
investigation of nulling for many pulsars in several works (e.g. \cite{rit76}
, \cite{ran86}, \cite{big92a}), has established that nulling is intrinsic 
to individual pulsars and possibly broadband in radio frequency.  
Though nulling is known to occur randomly, there are some recent studies by \cite{rw08} and 
\cite{hjr07} which report periodicity in nulling for quite a few pulsars. It is 
generally believed that onset of nulling is abrupt, i.e. pulse intensity drops
suddenly at the onset of a null. However, for PSR B0809$+$74 and PSR B1944$+$17, 
it is reported that the transitions from bursts to nulls show a gradual decline
of pulse energy, while transitions from nulls to bursts are abrupt (\cite{la83}, 
\cite{dchr86}). On the other hand, for PSR B0031$-$07, the onset of nulls is found 
to be abrupt \citep{viv95}. \cite{la83} investigated the pre and post null behavior 
of PSR B0809$+$74 and reported a relative dimness of the last few pulses before 
nulls, whereas the first active pulse at the onset of the burst (after each null)
appears to outshine the normal pulses. \cite{vsrr03} confirmed their results and 
also investigated the interaction between drifting and nulling
\begin{figure*}
\hbox{
 \hspace{1cm}
  \includegraphics[angle=0, width=0.42\textwidth]{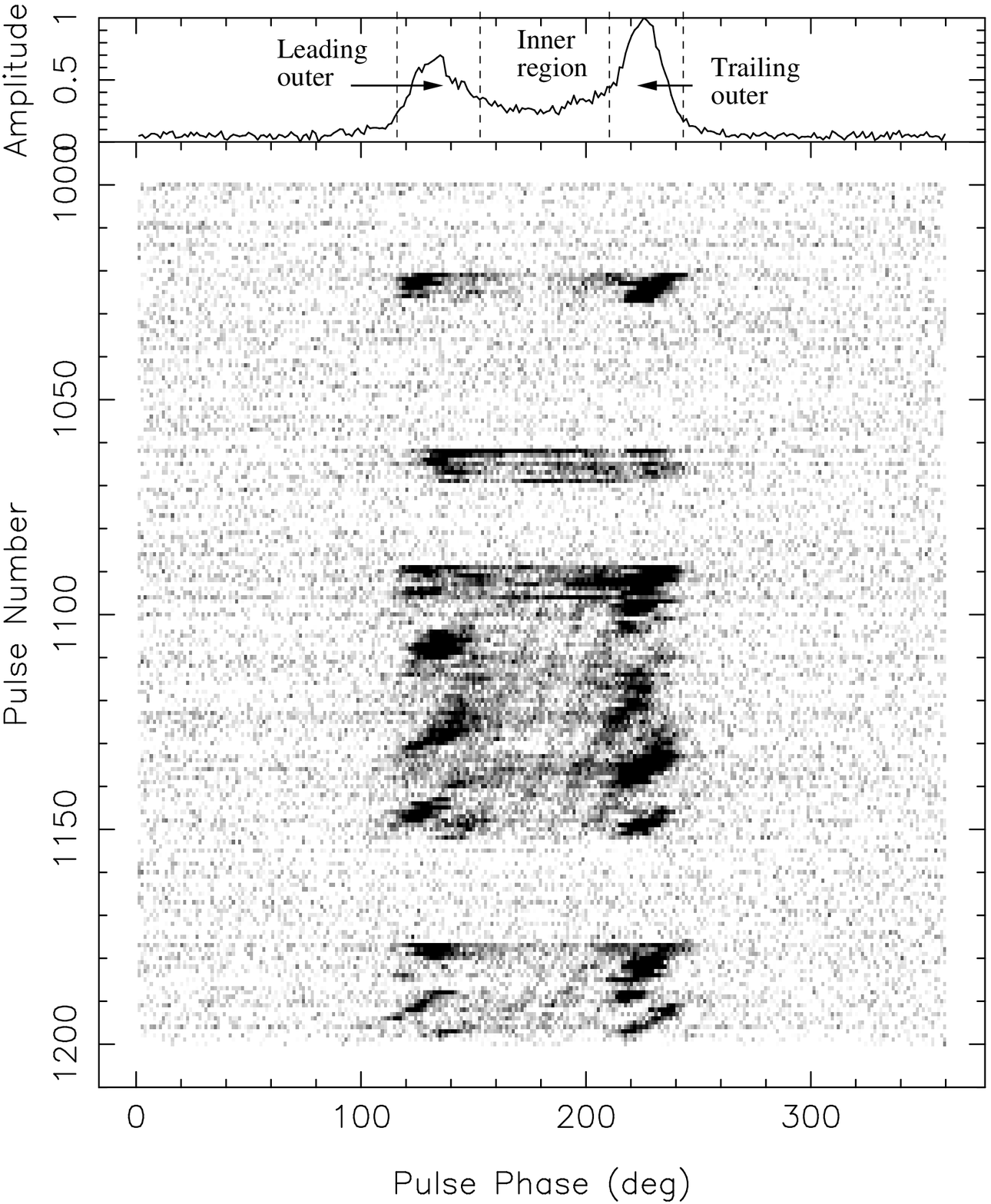}
  \includegraphics[angle=0, width=0.42\textwidth]{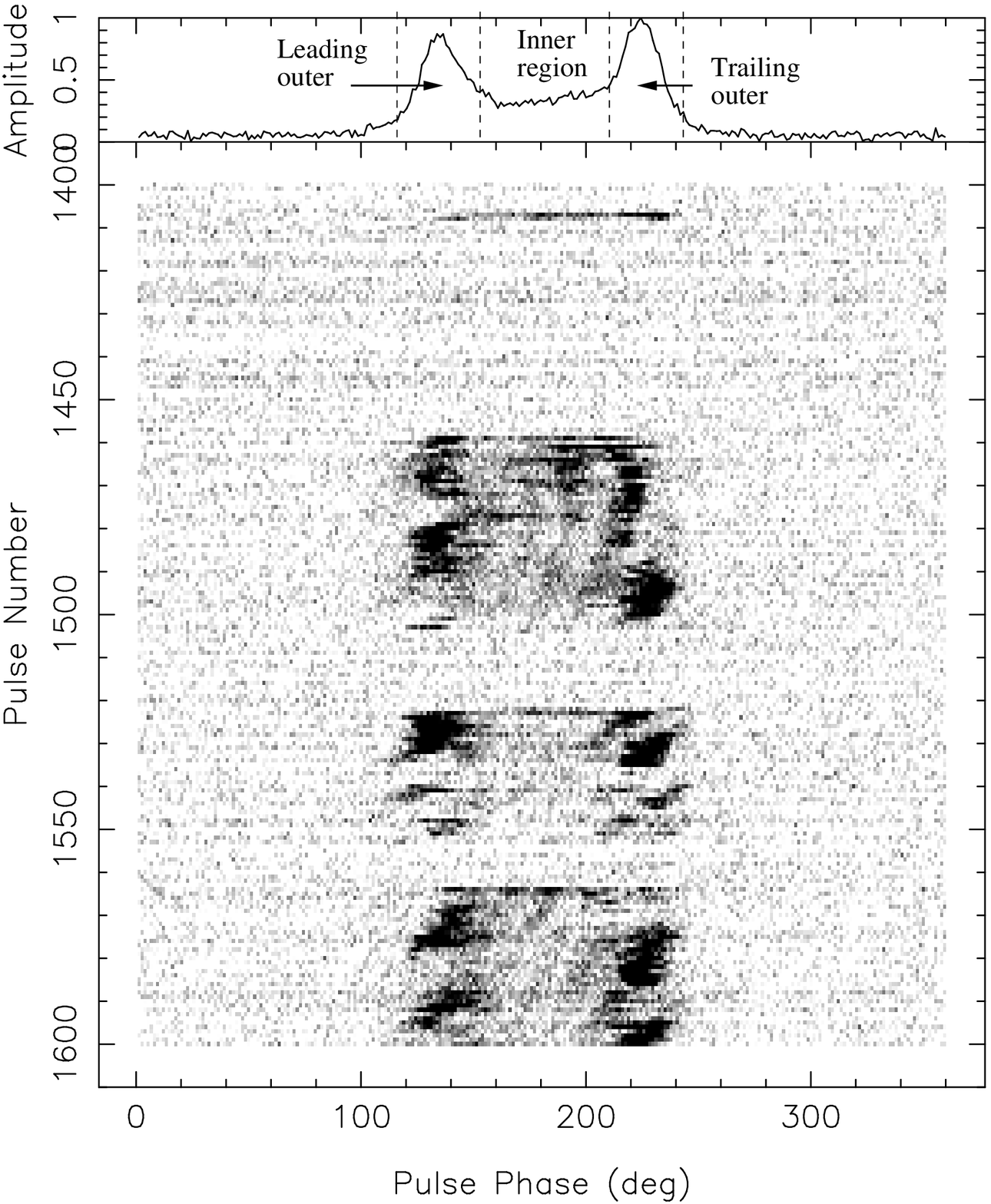}
   }
\caption[Gray scale plot of single pulses from PSR B0818$-$41 for pulse \#
1000 to 1200 and for pulse \# 1400 to 1600 from observations on 24 February
2004 at 325 MHz.]{Gray scale plot of single pulses from PSR B0818$-$41 from
observations on 24 February 2004 at 325 MHz. Left panel: pulse \# 1000
to 1200; pulsar nulls from pulse \# 1000 to 1020, pulse \# 1028 to 1061,
pulse \# 1070 to 1088 and pulse \# 1153 to 1176. Right panel: pulse \#
1400 to 1600; pulsar nulls from pulse \# 1409 to 1458, pulse \# 1504 to 1521
and pulse \# 1552 to 1563.}
\label{ch4_sp1_325}
\end{figure*}
in PSR B0809$+$74. They concluded that the drift pattern immediately after
the nulls differs from the normal one and commented that, beside its normal
and most common mode, the pulsar emits in a significantly different
quasi-stable mode immediately after most, or possibly all, the nulls. In this
mode the pulsar is brighter and the subpulse separation is less. They also
reported that the subpulses drift more slowly and the pulse window is
shifted towards the earlier longitudes. Investigating the interaction
between drifting and nulling, \cite{jv04} determined the alias order for
drifting of PSR B0818$-$13.

Though there has been significant progress both in the field of observations
as well as understanding and characterising the phenomenon of nulling, the
reason behind pulsar nulling and its connection to the emission mechanism is 
not tightly pinned down. In this regard, study of the emission properties 
before and after nulls, for individual pulsars, assumes importance. 
Remarkable subpulse drift pattern and frequent nulling is observed in PSR
B0818$-$41 \citep{Bhattacharyya_etal}. We find simultaneous occurrence of three 
drift regions with two different drift rates: an inner region with steeper 
apparent drift rate, flanked on each side by a region of slower apparent drift 
rate. 
The closely spaced drift bands always maintain a constant phase relationship: 
the subpulse emission from the inner drift region is in phase with that from 
the outer drift region on the right hand
side, and at the same time the emission in the inner drift region is out of
phase with the outer drift region situated on the left hand side. This phase
locked relationship is maintained for the entire stretch of the data (for
all the epochs of observations at 325 and 610 MHz) and does not appear to
get perturbed after intermittent nulling or during changes in the drift
rates. Although an extensive study of subpulse drifting and polarization
properties of this pulsar is presented in \cite{Bhattacharyya_etal_09}, its
nulling properties remain hitherto unexplored. This paper reports a detailed 
investigation of the behaviour of this pulsar around nulls, from observations
at 325 and 610 MHz. Results from our study are presented in Sect. \ref{ch4_analysis}. 
In Sect. \ref{ch4_discussion} we discuss about the implications of the results 
and explain these with Partially Screened Gap model \citep{Gil_etal_03}. Finally 
in Sect. \ref{sec:summary} we summarise our findings. 

\begin{center}
\begin{figure}
\begin{center}
\includegraphics[angle=0,width=0.42\textwidth]{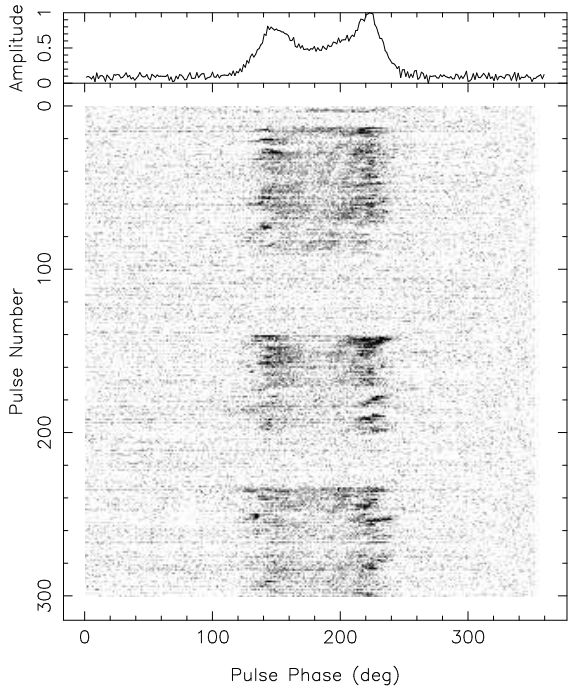} 
\vspace{0.2cm}
\end{center}
\caption[Same as Fig. \protect\ref{ch4_sp1_325}, but for pulse \# 1 to 300
from observations on 25 February 2004 at 610 MHz]{Same as Fig. \protect\ref%
{ch4_sp1_325}, but for pulse \# 1 to 300 from observations on 25 February
2004 at 610 MHz. Pulsar nulls from pulse \# 3 to 13, pulse \# 92 to 141 and
pulse \# 202 to 234.}
\label{ch4_sp_610}
\end{figure}
\end{center}


\begin{center}
\begin{figure}
\begin{center}
\includegraphics[angle=0,width=0.5\textwidth]{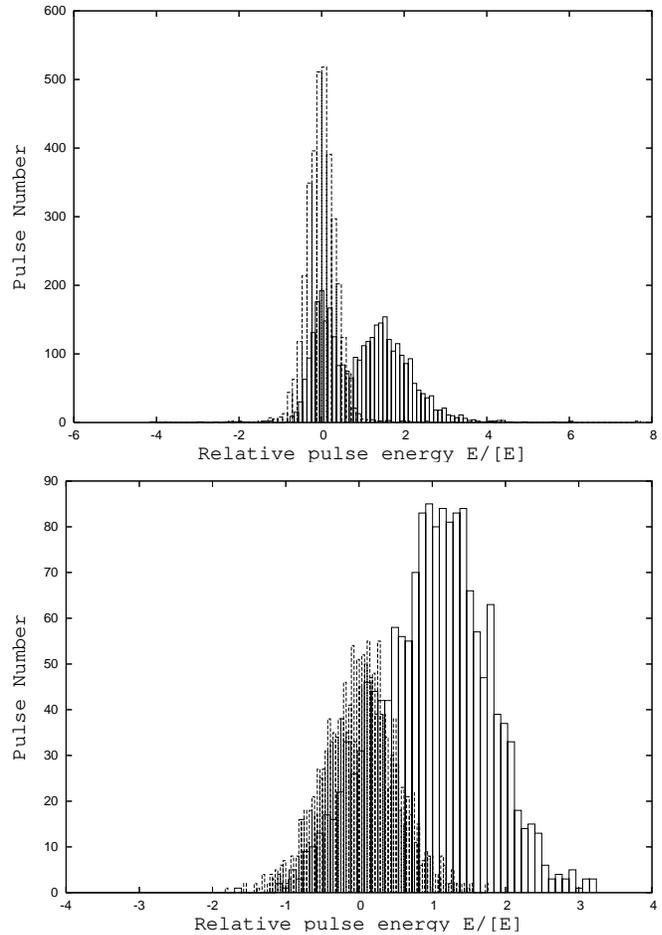}
\end{center}
\caption[Energy histogram for PSR B0818$-$41 at 325 and 610 MHz]{Energy
histograms for PSR B0818$-$41. Those for the on pulse window are bars with
solid lines; off pulse window are bars with dotted lines.  Top panel: at 
325 MHz with 3414 pulses, from observations on 24 February 2004. Bottom 
panel: at 610 MHz with 1612 pulses, from observations on 25 February 2004.}
\label{hist_325_610}
\end{figure}
\end{center}

\section{Analysis and Results}

\label{ch4_analysis} For the investigation of nulling behavior of PSR B0818$%
- $41, we have used single pulse observations at 325 MHz for two epochs (24 
February 2004 and 21 December 2005, containing 3414 and 6600 pulses 
respectively) and at 610 MHz for two epochs (25 February 2004 and 11 January 
2005, containing 1612 and 3600 pulses respectively). Details of the
observations are described in Table 1 of \cite{Bhattacharyya_etal_09}.

We observe frequent nulling for PSR B0818$-$41. Duration of the nulls varies
from few tens of pulses to a maximum of about a few hundred pulses. Fig. 
\ref{ch4_sp1_325} shows a sample of the single pulse gray scale plots of 
PSR B0818$-$41 with both drifting and nulling at 325 MHz, from the GMRT 
observations on 24 February 2004. Fig. \ref{ch4_sp_610} plots the same, but 
from observations on 25 February 2004 at 610 MHz. Instances of pulsar in 
the null state are mentioned in the captions. Investigating the single pulse 
gray scale plots, we observe that the first active pulses in the bursts after 
the nulls look different from the normal pulses. For these pulses, the inner 
region appears more filled and is significantly more intense than that for 
the normal pulses (e.g. pulse \#s 1062, 1089, 1177, 1459, 1522 and 1564 in Fig. \ref{ch4_sp1_325}). 
Though pulse \# 1021, occurs just after a null and does not greatly outshine 
the following pulses, it is brighter than the following pulses and is significantly 
brighter than the last active pulses before the null. Such cases with relatively 
less intense active pulses just after the nulls are rare, seen 
for $\sim$ 5\% of all the nulls, associated with shorter nulls (<10 pulses).
On the other hand, the inner regions of the pulses just before the onset of 
the nulls (e.g. pulse \#s 1027, 1152, 1503, 1551 in Fig. \ref{ch4_sp1_325}) appear 
less bright. In spite of being just before the onset of a null, pulse \# 1069 is not 
less intense; this is also a rare exception seen typically for shorter nulls 
(<10 pulses), and consists $\sim$ 3\% of all the nulls. Association of changing 
drift rates with the nulls, which is discussed in details in Sect. 7.1 of 
\cite{Bhattacharyya_etal_09}, is quite evident in the single pulse gray scale plots.

\subsection{Null distribution and identification of active and null pulses}

Identification of the active and null states depends on the sensitivity
limit of the telescope. High signal to noise (S/N) single pulse observations 
are required for accurate characterization of pulsar nulling properties. 
\cite{rit76} and \cite{big92a} investigated the statistics of pulse energy
distributions for characterizing the phenomenon of nulling. We follow a
similar procedure for PSR B0818$-$41 and the corresponding ON pulse as well
as OFF pulse energy histograms at 325 MHz from the observations on 24
February 2004, are shown in Fig. \ref{hist_325_610}. 
At 325 MHz, the ON pulse histogram has two components corresponding to active
and null pulses. We note (i) the strong presence of pulses with zero or near
zero aggregate intensity, clearly indicating the occurrence of nulling and 
(ii) that the distribution is continuous between the pulses and the nulls, 
making it difficult to separate out the populations. At 610 MHz, the 
distributions tend to merge with each other as a consequence of lower S/N 
(Fig. \ref{hist_325_610}). Using the method described in \cite{rit76}, we 
calculate a nulling fraction of $\sim$30\% at 325 MHz, from the data of 24 
February 2004, and a matching value of $\sim$28\% from the data of 21 
December 2005. We can not estimate meaningful values from the 610 MHz data, 
as the ON and OFF distributions overlap significantly at this frequency, but the nulling 
fraction at 610 MHz seems to be consistent with the 325 MHz data.

\begin{center}
\begin{figure*}
\begin{center}
\includegraphics[angle=0,width=1\textwidth]{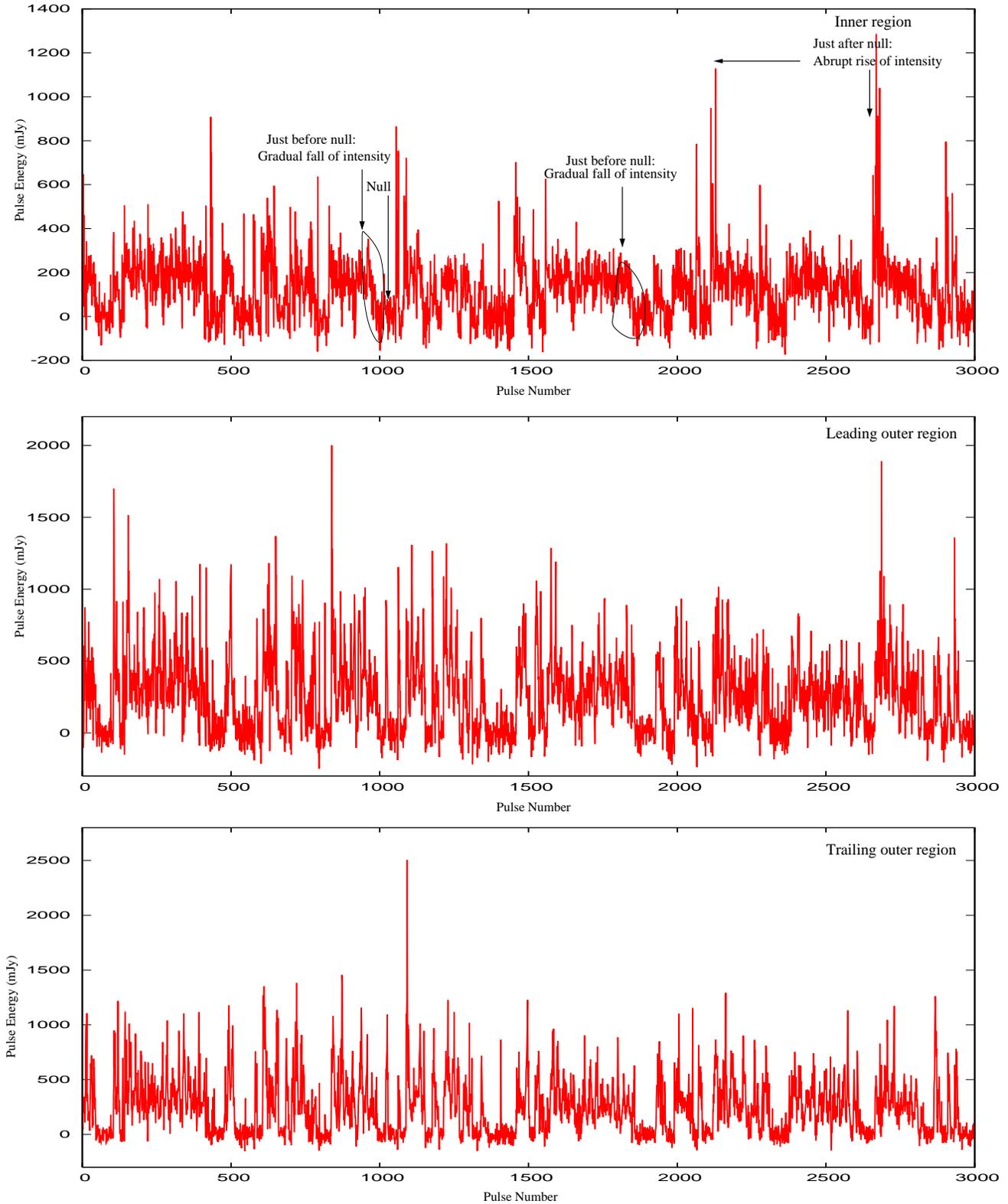}
\end{center}
\caption[]{Top panel: Total energy of the inner region versus pulse number 
for the 3000 pulses of the 325 MHz data of 24 February 2004.  Middle panel: 
Same as top panel, but for the leading outer region. Bottom panel: Same as 
top panel, but for the trailing outer region. Marked in the top panel are
some instances of pulsar nulling, and the gradual decrease of intensity 
before the onset of the null, as well as examples of the sudden increase
of intensity in the first few pulses when the pulsar turns on after the 
end of the null.}
\label{ch4_pe_325}
\end{figure*}
\end{center}

Although pulse energy distributions provide statistical information about
the nulling phenomenon, identifying individual nulls at each frequency is
required for more detailed investigation. This is done by comparing the ON
pulse energy estimate with a threshold based on the system noise level. The
uncertainty in the pulse energy estimate $\sigma_{ep, on}$ is given by $%
\sqrt{n_{on}}\sigma_{off}$, where $n_{on}$ is the number of ON pulse bins
and $\sigma_{off}$ is the rms of the OFF pulse region. Using this as a
threshold, we classify pulses with ON pulse energy smaller than $3
\times\sigma_{ep, on}$, as null pulses.  Using this, we are able to 
construct a ON/OFF time series of the single pulses, identify the lengths
of the individual nulls and bursts, and also sequence the ON pulses just 
before a null and those in the burst just after each null.  

\subsection{Durations of nulls and bursts}
We aim to investigate any possible connection between the lengths of the 
bursts and the nulls. For example, does waiting longer for a null mean that 
it will last longer too? The durations of the nulls and the bursts immediately 
before and after the nulls are tabulated in columns 2,3 and 4 of Tables
A1, A2, A3 and A4.
We do not see any obvious correlation between the 
lengths of the nulls and the corresponding bursts, as
well as between the null or burst lengths and the relative strengths of the
pulses just before and after the nulls. Our result is similar to the study
of PSR B0809$+$74 by \cite{vsrr03}, who found that the lengths of the
neighboring nulls and bursts are independent. This may indicate that there
is no systematic dependence between the mechanism for nulling and the
duration of nulls and bursts.


\begin{center}
\begin{table*}
\begin{minipage}{170mm}
\caption[Investigation of the intensity distribution of the inner region immediately before and after the nulls
(from observations at 325 MHz on 24 February 2004)]{Investigation of the intensity distribution of the inner region
immediately before and after the nulls (from observations at 325 MHz on 24 February 2004)}
\vspace{0.3cm}
\label{ch4_table:null_1}
\begin{tabular}{|c|c|c|c|c|c|c|c|c|c|c|c|c|c|c|c|c|}
\hline
Serial  & Duration & Duration & Duration & $\frac{I_{a}}{\langle I_{b}(1p:10p)\rangle}$& $\frac{\langle I_{a}(4p:10p)\rangle}{\langle I_{b}(1p:10p)\rangle}$ & $\frac{I_{a}}{\langle I_{a}(4p:10p)\rangle}$ \\
Number  & of null  & of burst & of burst &       &       &       \\
        &          & after    & before   &       &       &       \\
        &          & null     & null     &       &       &       \\\hline
  1     & 59       & 20       & $>$46      & 1.89  & 0.86  & 2.19  \\
  2     & 5        & 276      &  21      & 3.12  & 1.94  & 1.61 \\
  3     & 50       & 33       &  282     & 1.73  & 0.62  & 2.81 \\
  4     & 92       & 52       &  39      & 2.11  & 1.73  & 1.21 \\
  5     & 25       & 10       &  54      & 1.78  & 1.46  & 1.21 \\
  6     & 49       & 83       &  83      & 2.93  & 1.87  & 1.57 \\
  7     & 27       & 10       &  35      & 1.85  & 0.61  & 3.01 \\
  8     & 33       & 10       &  10      & 5.97  & 3.43  & 3.12 \\
  9     & 18       & 64       &  10      & 1.28  & 1.12  & 1.15 \\
  10    & 23       & 21       &  64      & 1.83  & 0.71  & 2.55 \\
  11    & 14       & 52       &  21      & 2.07  & 1.52  & 1.35 \\
  12    & 12       & 31       &  52      & 1.02  & 0.69  & 1.46 \\
  13    & 28       & 17       &  31      & 1.49  & 0.85  & 1.76 \\
  14    & 17       & 31       &  40      & 1.67  & 0.99  & 1.69 \\
  15    & 11       & 294      &  31      & 5.56  & 1.89  & 2.95 \\
  16    & 17       & 14       &  64      & 5.52  & 2.70  & 2.04 \\
  17    & 34       & 150      &  14      & 3.07  & 1.47  & 2.09 \\
  18    & 8        & 28       &  150     & 6.99  & 2.36  & 2.97 \\
  19    & 62       & 207      &  28      & 1.43  & 0.62  & 2.28 \\
  20    & 10       & 35       &  207     & 2.11  & 0.91  & 2.32 \\
  21    & 37       & 148      &  35      & 5.20  & 3.01  & 1.73 \\
  22    & 10       & 11       &  10      & 1.75  & 0.87  & 2.02 \\\hline
Mean value &  $-$  & $-$      &  $-$     & 2.84  & 1.46  & 2.05 \\\hline
Median Value& $-$  & $-$      &  $-$     & 1.98  & 1.29  & 2.03 \\\hline
\end{tabular}
\end{minipage}
\end{table*}
\end{center}


\begin{center}
\begin{table*}
\begin{minipage}{170mm}
\caption[Same as Table \ref{ch4_table:null_1}, but from another epoch at 325 MHz on 21 December 2005]{Same as Table \ref{ch4_table:null_1}, but from another epoch at 325 MHz on 21 December 2005}
\vspace{0.3cm}
\label{ch4_table:null_2}
\begin{tabular}{|c|c|c|c|c|c|c|c|c|c|c|c|c|c|c|c|}
\hline
Serial  & Duration & Duration & Duration &$\frac{I_{a}}{\langle I_{b}(1p:10p)\rangle}$& $\frac{\langle I_{a}(4p:10p)\rangle}{\langle I_{b}(1p:10p)\rangle}$ & $\frac{I_{a}}{\langle I_{a}(4p:10p)\rangle}$ \\
Number  & of null  & of burst & of burst &        &       &                                    \\
        &          & after    & before    &        &       &                                     \\
        &          & null     & null     &        &       &                                     \\\hline
  1     & 46       &  30      & $>$46      & 2.25   & 1.67  & 1.53                                \\
  2     & 83       &  203     & 474      & 5.95   & 2.10  & 2.84                                \\
  3     & 28       &  202     & 185      & 2.18   & 1.02  & 2.15                                \\
  4     & 10       &  250     & 202      & 2.25   & 1.31  & 1.71                                \\
  5     & 268      &  145     & 103      & 6.83   & 1.96  & 3.48                               \\
  6     & 86       &  150     & 145      & 9.28   & 2.20  & 4.21                                \\
  7     & 22       &  51      & 150      & 2.91   & 1.38  & 2.11                                \\
  8     & 10       &  125     & 56       & 2.39   & 1.83  & 1.31                               \\
  9     & 104      &  47      & 120      & 1.84   & 0.55  & 3.35                               \\
  10    & 48       &  49      & 274      & 2.34   & 0.80  & 2.90                                \\
  11    & 23       &  17      & 49       & 2.38   & 1.08  & 2.20                                \\
  12    & 50       &  105     & 17       & 1.87   & 1.24  & 1.50                               \\
  13    & 41       &  15      & 370      & 1.82   & 1.84  & 0.99                                \\
  14    & 330      &  109     & 15       & 1.01   & 0.71  & 1.42                               \\
  15    & 8        &  50      & 109      & 1.26   & 0.96  & 1.30                               \\
  16    & 159      &  184     & 50       & 3.06   & 2.33  & 1.31                                \\
  17    & 23       &  22      & 184      & 1.56   & 0.93  & 1.67                                \\
  18    & 103      &  33      & 199      & 4.39   & 2.02  & 2.18                               \\
  19    & 22       &  73      & 30       & 1.67   & 0.76  & 2.19                               \\\hline
Mean value &  $-$  &  $-$     & $-$      & 3.03   & 1.40  & 2.12                                \\\hline
Median value& $-$  &  $-$     & $-$      & 2.33   & 1.31  & 2.11                                \\\hline
\end{tabular}
\end{minipage}
\end{table*}
\end{center}


\begin{center}
\begin{table*}
\begin{minipage}{170mm}
\caption[Same as Table \ref{ch4_table:null_1}, but at 610 MHz on 25 February 2004]{Same as Table \ref{ch4_table:null_1}, but at 610 MHz on 25 February 2004}
\vspace{0.3cm}
\label{ch4_table:null_3}
\begin{tabular}{|c|c|c|c|c|c|c|c|c|c|c|c|c|c|c|c|}
\hline
 Serial  & Duration & Duration & Duration &$\frac{I_{a}}{\langle I_{b}(1p:10p)\rangle}$& $\frac{\langle I_{a}(4p:10p)\rangle}{\langle I_{b}(1p:10p)\rangle}$ & $\frac{I_{a}}{\langle I_{a}(4p:10p)\rangle}$ \\
 Number  & of null  & of burst & of burst &       &      &                                      \\
         &          & after    & before   &       &      &                                      \\
         &          & null     & null     &       &      &                                      \\\hline
   1     & 8        &   81     & $>$5       &  1.88 & 0.52 & 3.64                                \\
   2     & 47       &   61     & 81       &  2.15 & 0.76 & 2.85                                \\
   3     & 32       &   793    & 61       &  3.32 & 2.59 & 1.28                                \\
   4     & 98       &   24     & 793      &  3.19 & 1.19 & 2.66                                \\
   5     & 33       &   78     & 24       &  1.53 & 0.97 & 1.58                                \\
   6     & 28       &   15     & 78       &  2.44 & 0.81 & 3.02                                \\
   7     & 19       &   48     & 15       &  1.51 & 1.25 & 1.21                                 \\
   8     & 37       &   47     & 48       &  2.29 & 1.73 & 1.33                                \\
   9     & 18       &   115    & 47       &  2.26 & 1.81 & 1.24                                \\\hline
Mean value &  $-$   &   $-$    & $-$      &  2.29 & 1.43 & 1.81                                 \\\hline
Median value & $-$  &   $-$    & $-$      &  2.26 & 1.35 & 1.36                                 \\\hline
\end{tabular}
\end{minipage}
\end{table*}
\end{center}


\begin{center}
\begin{table*}
\begin{minipage}{170mm}
\caption[Same as Table \ref{ch4_table:null_1}, but at 610 MHz on 11 January 2005]{Same as Table \ref{ch4_table:null_1}, but at 610 MHz on 11 January 2005}
\vspace{0.3cm}
\label{ch4_table:null_4}
\begin{tabular}{|c|c|c|c|c|c|c|c|c|c|c|c|c|c|c|c|}
\hline
 Serial  & Duration & Duration & Duration &$\frac{I_{a}}{\langle I_{b}(1p:10p)\rangle}$& $\frac{\langle I_{a}(4p:10p)\rangle}{\langle I_{b}(1p:10p)\rangle}$ & $\frac{I_{a}}{\langle I_{a}(4p:10p)\rangle}$ \\
 Number  & of null  & of burst & of burst &       &      &                                      \\
         &          & after    & before   &       &      &                                      \\
         &          & null     & null     &       &      &                                      \\\hline
   1     & 200      &  400     & 725      &  2.51 & 1.70 & 1.47                                \\
   2     & 60       &  113     & 96       &  2.95 & 3.93 & 0.75                                \\
   3     & 73       &  55      & 113      &  1.42 & 1.15 & 1.24                                \\
   4     & 31       &  164     & 55       &  2.51 & 2.13 & 1.18                                \\
   5     & 86       &  130     & 164      &  2.88 & 2.42 & 1.19                                \\
   6     & 40       &   15     & 130      &  1.80 & 0.84 & 2.13                                \\
   7     & 94       &  220     & 405      &  0.90 & 1.03 & 0.87                                 \\
   8     & 128      &  103     & 202      &  2.47 & 1.26 & 1.96                                \\\hline
Mean value &   $-$  &  $-$     & $-$      &  2.18 & 1.81 & 1.35                                 \\\hline
Median value & $-$  &  $-$     & $-$      &  2.49 & 1.48 & 1.21                                 \\\hline
\end{tabular}
\end{minipage}
\end{table*}
\end{center}



\begin{center}
\begin{figure}
\begin{center}
\includegraphics[angle=0,width=0.5\textwidth]{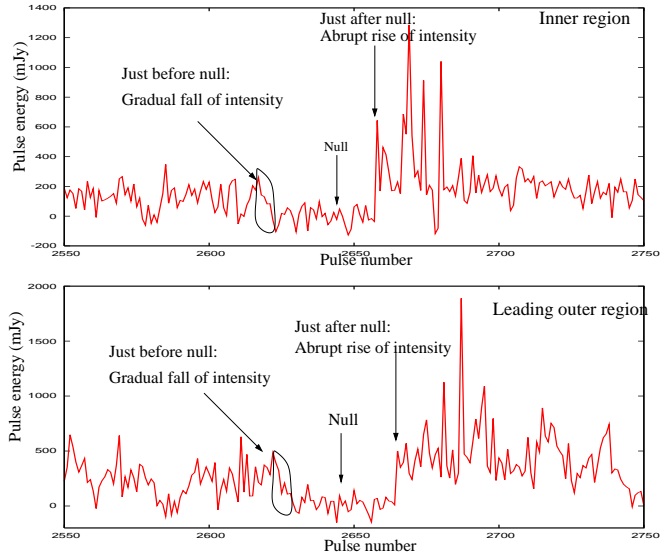}
\end{center}
\caption[]{Top panel: Total energy of the inner region versus pulse number 
for a zoom in on pulse \#s 2400 to 2800 from the 325 MHz observations on 
24 February 2004. Bottom panel: Same as top panel, but for the leading outer 
region.}
\label{ch4_pe_325_zoom}
\end{figure}
\end{center}


\subsection{Pulse intensity variations before and after individual nulls} \label{ch4_int_indnull} 
Fig. \ref{ch4_pe_325} plots the pulse energy versus the pulse number 
of the inner region as well as the leading and trailing outer regions
of the pulse profile, for pulse \# 1 to 3000 from observations on 24 
February 2004 at 325 MHz (see Fig. \ref{ch4_sp1_325} for definition of 
the inner and outer drift regions).  The nulls are identified, in these
plots, as the bunches of pulses of low energy, close to zero.  An 
interesting trend can be seen, which is most prominent for the inner 
region (top panel of the figure) -- the pulse intensity gradually goes 
down before the onset of a null, and suddenly shoots up for the first 
few active pulses in the burst. This trend is also present in 
leading and trailing outer regions, but is not as clear.  A zoom in
on one such typical event is shown in Fig. \ref{ch4_pe_325_zoom}, 
which plots the pulse energy for the inner and leading outer regions
for pulse \# 2550 to 2750.  The gradual switching off before the null
starts and the abrupt switching on at the beginning of the burst are
clearly seen for both the emission regions, though with some differences
in detailed properties, which are explored further in Sect. \ref{ch4_avp}. 

From these results we infer that the entire pulse energy appears to change
in a fairly particular manner immediately before and after the nulls, though
the variations are more clear for the inner region of the pulse.  We believe
this difference in behaviour may be due to the fact that the intensities of 
the leading and trailing outer regions of successive pulses are already 
significantly modulated at the 18.3 $P_1$ periodicity (referred to as $P_3^m$,
see \cite{Bhattacharyya_etal}), making it harder to detect the variations
we are looking for.
For the inner region, the averaging over the multiple drift bands helps to
smooth out the $P_3^m$ modulation effect.
This is a reflection of the fact that in case of inner region our line of sight
is grazing the emission ring, whereas for the leading and trailing outer regions 
we have a more central or direct traverse  (see Fig. 7 and Fig. 8 of 
\cite{Bhattacharyya_etal_09}). 

Consequently, we choose the inner region for further quantitative investigation 
of these intensity modulations immediately before and after the nulls. We select 
the sequences of nulls for which pulsar is active for at least 10 pulses 
before the nulls, and the successive burst state lasts for at least 10 pulses.
For the data at 325 MHz, where we are able to identify a total of 41 such 
sequences from the 10,000 pulse data at the two epochs, we find that the time 
taken for transition from the active to null state varies somewhat, with a minimum of 
about 5 pulse periods, maximum of about 13 periods, and a mean of about 10 
periods. The transition from the null to the burst state is rather abrupt 
$-$ the intensity rises to a maximum over a short time scale, less than one 
period, which is maintained for a few pulses before decaying down to the
typical active state level.  A similar trend is seen in the 17 events that
are identified in the 5200 pulse data from the two epochs at 610 MHz. 

\begin{figure*}
\hbox{
  \includegraphics[angle=-90, width=0.51\textwidth]{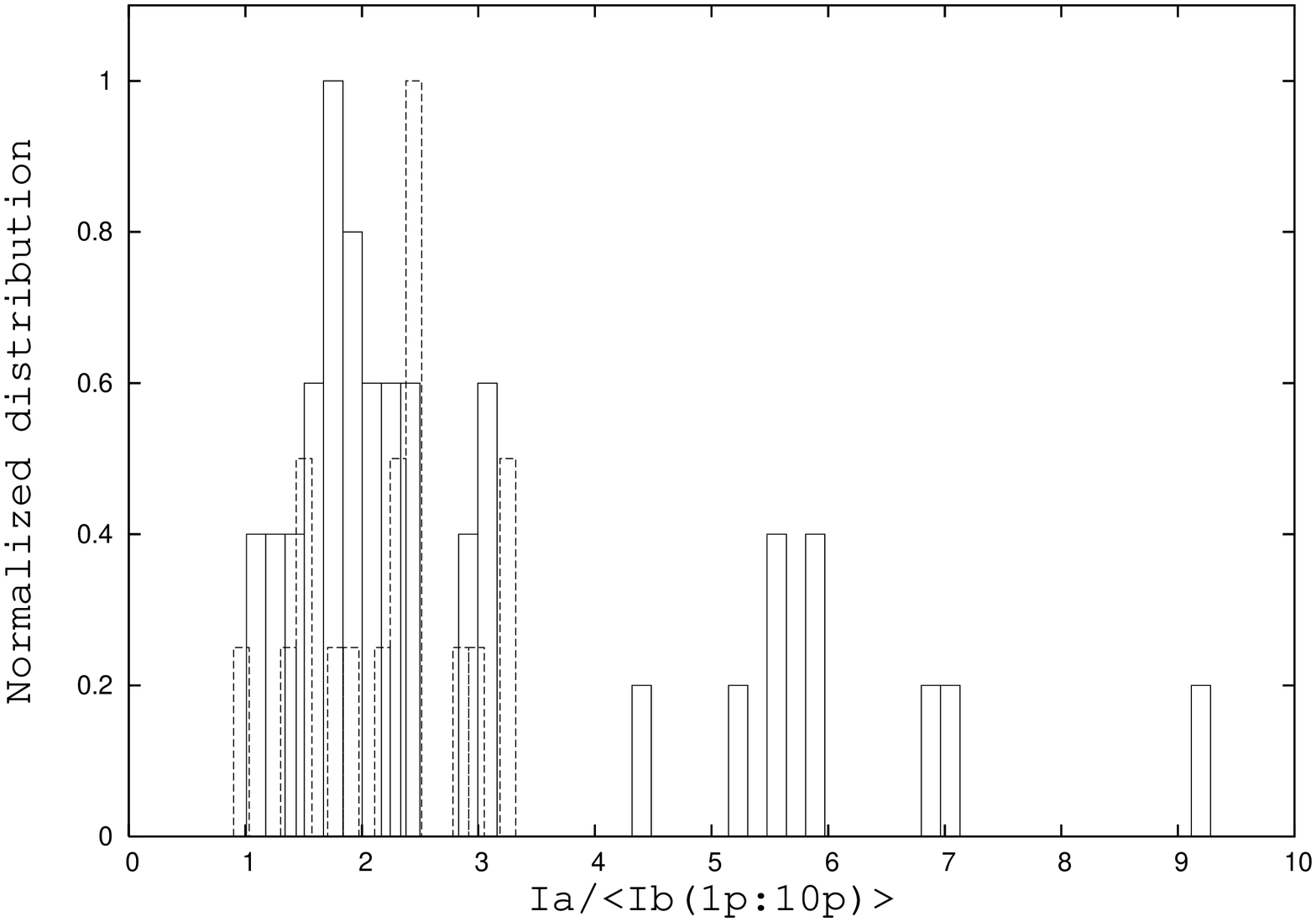}
  \includegraphics[angle=-90, width=0.51\textwidth]{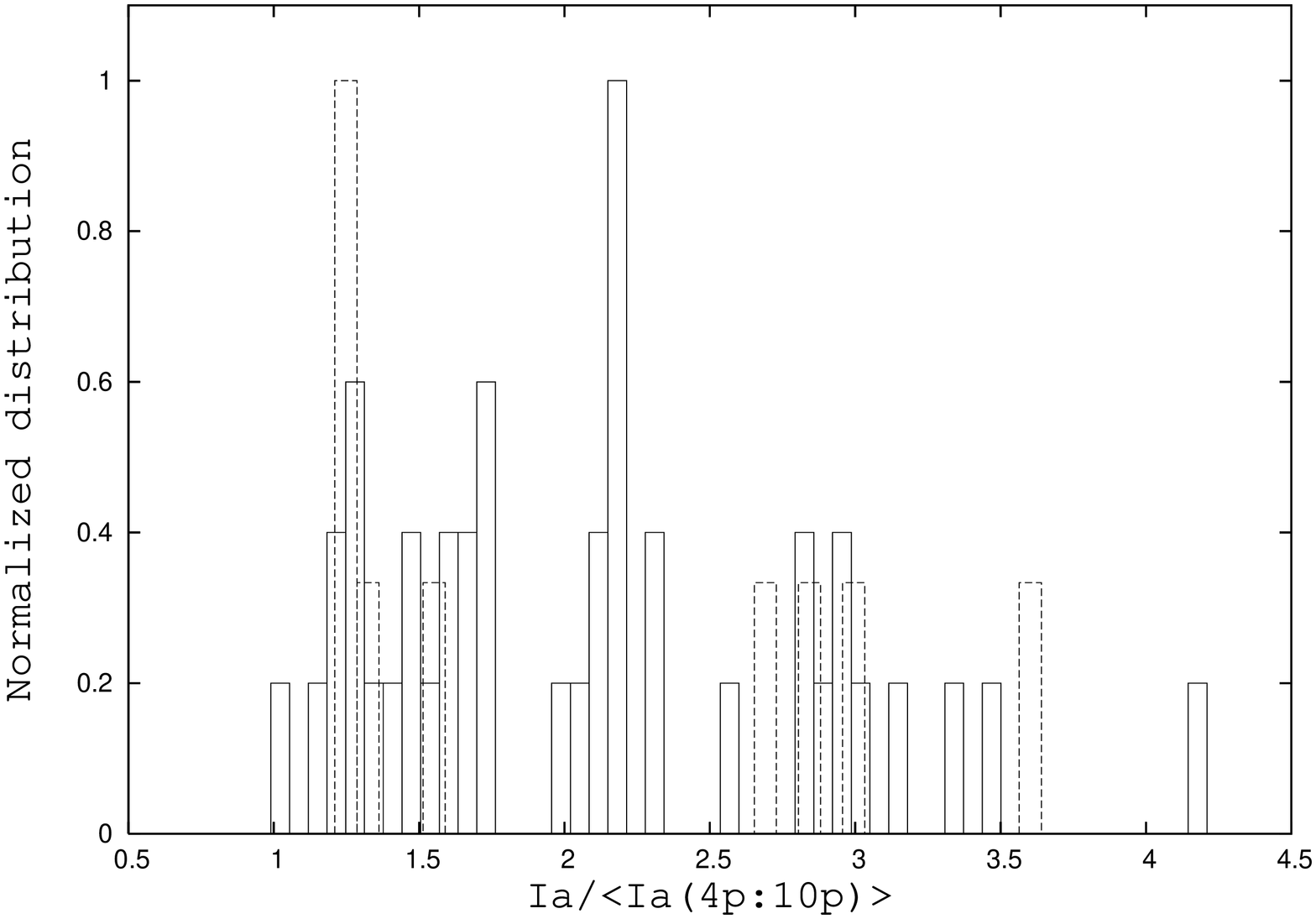}
}
\caption[Relative strength of the first active pulse just after the nulls to
the pulses before and after the nulls plotted against the null occurrence
number]{Relative strength of the first active pulse in the burst to
the pulses just before the onset of the null ($I_{a}/\langle I_{b}(1p:10p)\rangle$, 
shown in the left panel) and to the later pulses in the burst ($I_{a}/\langle I_{a}(4p:10p)\rangle$, 
shown in the right panel), from the ratios listed in Table \protect\ref{ch4_table:null_1}, 
\protect\ref{ch4_table:null_2}, \protect\ref{ch4_table:null_3} and \protect\ref{ch4_table:null_4}. 
The solid and dashed lines denote the corresponding 325 and 610 MHz values.}
\label{ch4:Ia_Ib10p}
\end{figure*}
\begin{figure*}
\hbox{
  \includegraphics[angle=-90, width=0.5\textwidth]{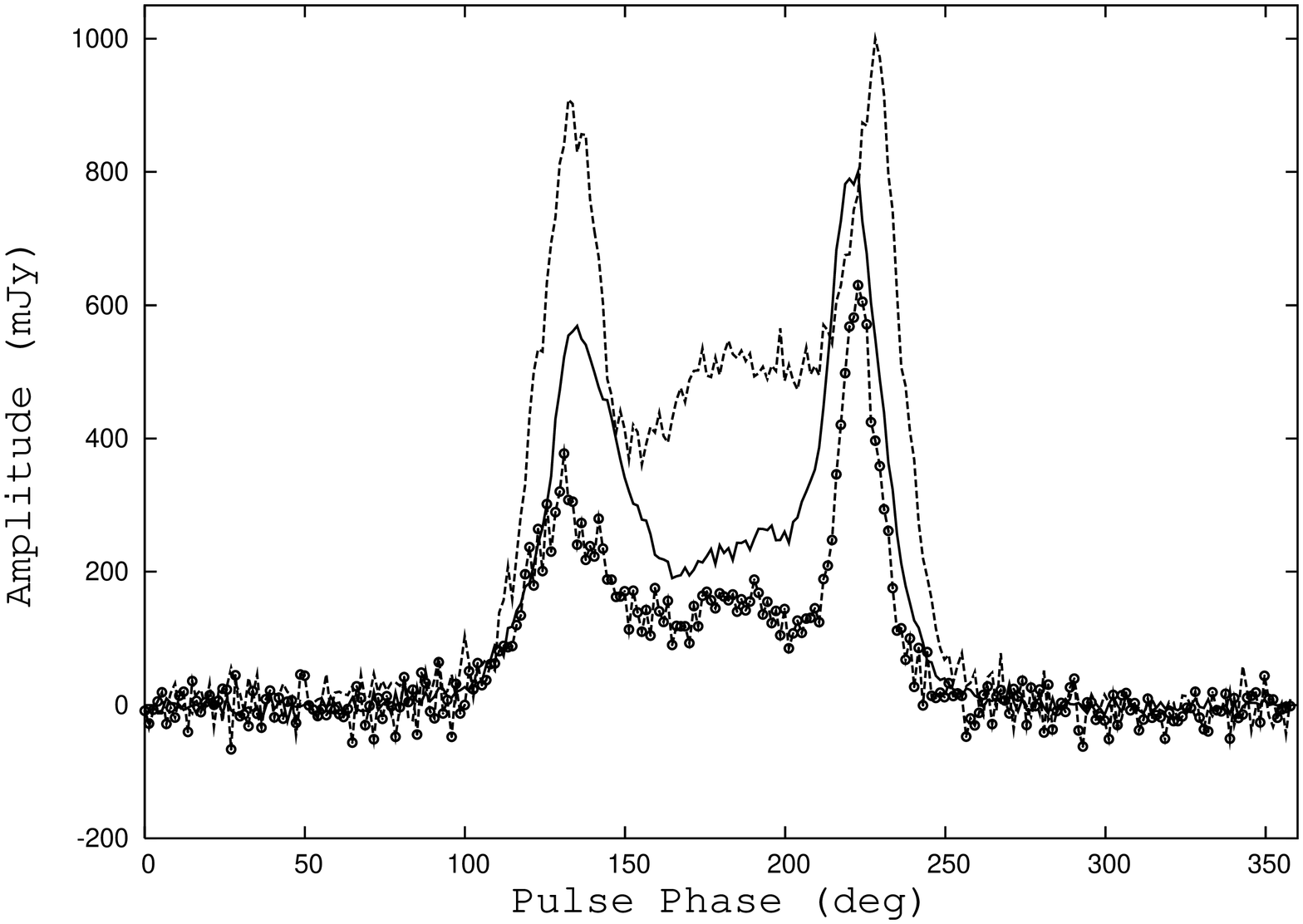}
  \includegraphics[angle=-90, width=0.5\textwidth]{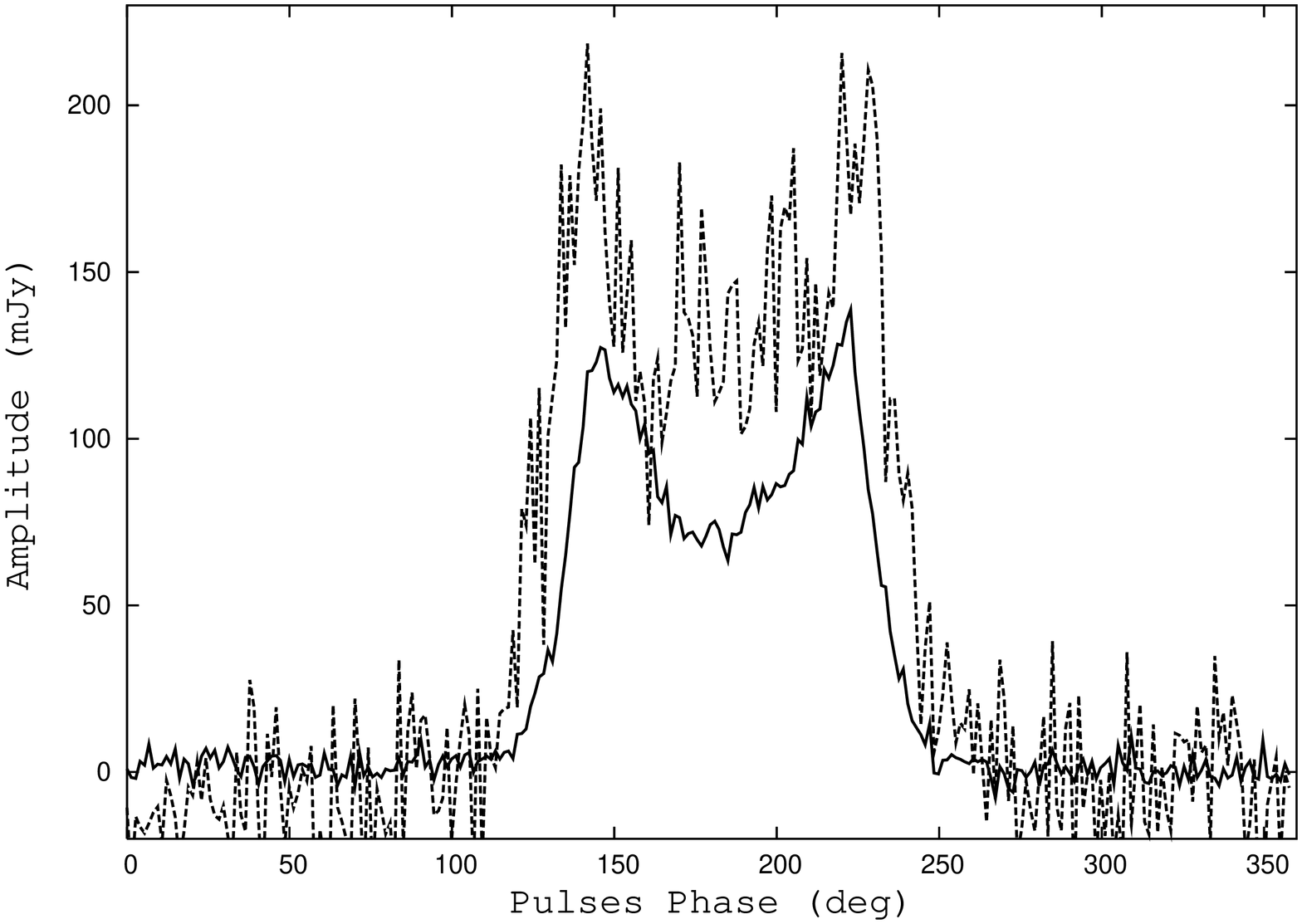}
}
\caption[Normal average profile and the average profile from the first
active pulses immediately after the nulls and last active pulses immediately
before the nulls at 325 and 610 MHz]{Left panel: The average profiles for regular
drifting mode (solid line), first active pulse of the bursts (dashed line),
last active pulses before the nulls (dashed line with open circles), at 325 
MHz from the data of 24 February 2004 and 21 December 2005. Right panel: 
Normal average profile (dashed line) and the average profile from the first 
active pulse of the bursts (solid line) at 610 MHz from the data of 25 February 
2004 and 11 January 2005.}
\label{ch4_avp325_24feb_21dec}
\end{figure*}

For each event, we compute the following quantities for the inner region,\newline
$\bullet$ $\langle I_{b}(1p:10p) \rangle$: mean intensity of 10 active
pulses immediately before the null.\newline
$\bullet$ $I_{a}$: intensity of the first active pulse of the burst just after
the null.\newline
$\bullet$ $\langle I_{a}(4p:10p)\rangle$: mean intensity of the 4th to 10th
active pulses in the burst. \footnote{The first three active pulses are 
generally stronger than the rest. So we considered the mean intensity of 
fourth pulse onwards for comparison of intensities just at the onset of the 
burst to sufficiently after the onset.}\newline
$\bullet$ $I_{a}/{\langle I_{b}(1p:10p)\rangle}$: ratio of $I_{a}$ to $
\langle I_{b}(1p:10p)\rangle$ which provides a comparison of the
strength of the first active pulse of the burst to the mean strength of 
the last ten active pulses before the null.\newline
$\bullet$ $I_{a}/{\langle I_{a}(4p:10p)\rangle}$: ratio of $I_{a}$ to $
\langle I_{a}(4p:10p)\rangle$ which provides a comparison of the
strength of the first active pulse of the burst to the mean strength 
further in the burst.\newline
$\bullet$ ${\langle I_{a}(4p:10p)\rangle}/{\langle I_{b}(1p:10p)\rangle}$:
ratio of $\langle I_{a}(4p:10p)\rangle$ to $\langle I_{b}(1p:10p)\rangle$
which provides a comparison of the strength of the active pulses in the 
burst (from the fourth pulse onwards and up to the tenth pulse) to the 
mean of the last ten active pulses before the null.

The three ratios above are tabulated in Table \ref{ch4_table:null_1} for 
the 325 MHz observations on 24 February 2004\footnote{Same exercise tried 
with $\langle I_{a}(1p:3p)\rangle$ or $\langle I_{a}(1p:2p)\rangle$ in place 
of $I_{a}$ produces similar ratios}.
We find that ${I_{a}}/{\langle I_{b}(1p:10p)\rangle}$ is always greater than 
unity with a mean of 2.8, confirming that first active pulse in the burst 
outshines the pulses just before the nulls. ${\langle I_{a}(4p:10p)\rangle}/{
\langle I_{b}(1p:10p)\rangle}$ is comparatively less and oscillates around
unity, with the mean over all the events coming to 1.5. 
${I_{a}}/{\langle I_{a}(4p:10p)\rangle}$ is always greater than unity (mean 
value $\sim$ 2.0) signifying that the first few active pulses in the burst
are brighter than those later in the burst. Table \ref{ch4_table:null_2} 
lists the same as Table \ref{ch4_table:null_1}, but for observations at 
325 MHz on the second epoch $-$ the results are found to be very similar for 
both the epochs.
Tables \ref{ch4_table:null_3} and \ref{ch4_table:null_4} give the corresponding 
results for 610 MHz, from data of 25 February 2004 and 11 January 2005. 
${I_{a}}/{\langle I_{b}(1p:10p)\rangle}$ is almost always more than unity 
with a mean around 2.2 at each epoch, which is a bit lower than the values 
at 325 MHz. Same is true for the comparison of ${I_{a}}/{\langle I_{a}(4p:10p)\rangle}$ 
values between 325 and 610 MHz. However, the mean value of $\langle I_{a}(4p:10p)\rangle/{\langle
I_{b}(1p:10p)\rangle}$ at 610 MHz is comparable or greater than the mean
values at 325 MHz. The mean of ${I_{a}}/{\langle I_{b}(1p:10p)\rangle}$ for all 
the null occurrences for both the frequencies is 2.8 and median is 2.3. Whereas the 
mean value of $I_{a}/\langle I_{a}(4p:10p)\rangle$ for all the null occurrences for 
both the frequencies is 2.0 and median is 1.9.

Left panel of Fig. \ref{ch4:Ia_Ib10p} is the plot of normalized distributions of 
${I_{a}}/{\langle I_{b}(1p:10p)\rangle}$ for all the epochs at 325 and 610 MHz. 
Right panel of Fig. \ref{ch4:Ia_Ib10p} plots the same for $I_{a}/\langle I_{a}(4p:10p)\rangle$.
The implications of the above findings are discussed in Sect. \ref{ch4_discussion}. 

From the above analysis we conclude that, at 325 MHz, the first few active 
pulses in the burst following a null are almost 3 times brighter than the 
last few pulses before the onset of the null (this ratio decreases to about 
2.2 at 610 MHz); also these pulses are about 2 times brighter than the 
successive pulse in the burst state (this ratio is again lower at 610 MHz).

\subsection{Average profile variations from pulses before and after nulls} \label{ch4_avp}

The analysis presented in the previous section brings out the fact that
the individual pulses immediately before and after the nulls have different
characteristics than the normal pulses. To study the mean behaviour, we 
construct the average profiles of the pulses immediately before and after
the nulls, for the nulls that satisfy the criterion described in Sect. \ref
{ch4_int_indnull}. The left panel of Fig. \ref{ch4_avp325_24feb_21dec} presents average pulse 
profiles of PSR B0818$-$41 at 325 MHz from the data of 24 February 2004 and 
21 December 2005 at 325 MHz from, (a) a sequence of 200 pulses showing regular 
drifting (hereafter referred to as normal profile) $-$ solid line, (b) the last 
active pulse before the nulls $-$ dashed line with open circles, and (c) from the 
first active pulse of the bursts $-$ dashed line. We can see clearly that the 
average profile from the last active pulse before the nulls is significantly 
weaker than the normal profile, for both the leading and trailing outer regions, as well 
as for the inner region.  It is also much weaker than the average profile from the
first active pulse in the bursts, which has a significant bump of enhanced
energy for the inner region, and comparable strengths for the leading and
trailing outer regions, quite unlike the other two profiles.
A similar behaviour is seen at 610 MHz. The right panel of Fig. \ref{ch4_avp325_24feb_21dec} 
presents the average pulse profiles of PSR B0818$-$41 at 610 MHz from the data of 
25 February 2004 and 11 January 2005, (a) a sequence of 300 pulses showing regular 
drifting (the normal average profile) $-$ solid line, and (b) average profile from the 
first active pulse of the bursts $-$ dashed line. The shapes of the profiles are somewhat 
similar, except for the bump of enhanced power in the inner region, which is also 
seen at 325 MHz. These results from average profiles at 325 and 610 MHz directly support
the conclusions obtained from the study of individual nulling events in
Sect. \ref{ch4_int_indnull}. It appears that the pulsar emerges from the
nulls very much brighter in overall intensity, with a very specific change
in pulse shape and energy distribution, which appears to be somewhat similar at the two frequencies.

\begin{center}
\begin{figure}
\begin{center}
\includegraphics[angle=-90,width=0.5\textwidth]{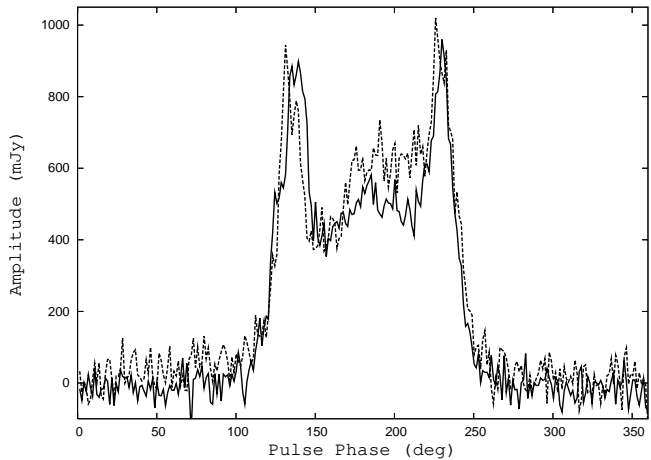}
\vspace{0.2cm}
\end{center}
\caption[Comparison of the average profile from first active pulses
immediately after the nulls at 325 MHz from observations on 24 February 2004
and on 21 December 2005]{Comparison of the average profile from the first
active pulse in the bursts from 325 MHz data of two epochs: 24 February 2004
(solid line) and 21 December 2005 (dashed line).}
\label{ch4_1p_aft_null}
\end{figure}
\end{center}



\begin{center}
\begin{figure}
\begin{center}
\includegraphics[angle=0,width=0.5\textwidth]{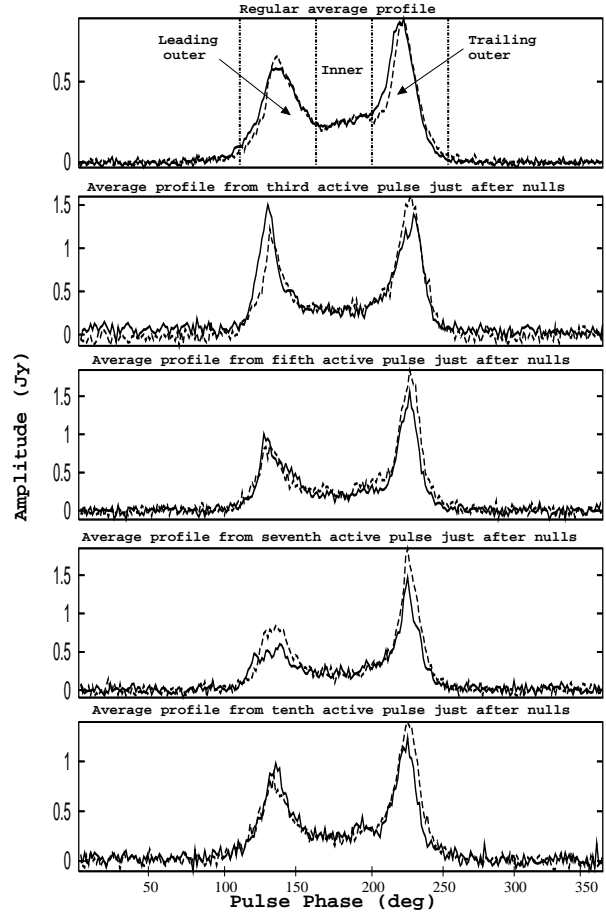}
\end{center}
\caption[Comparison of the average profile from the first few actives pulses
immediately after the nulls at 325 MHz from observations on 24 February 2004
and on 21 December 2005]{Comparison of the average profile from the third, fifth,
seventh and tenth pulse in the bursts, for the two epochs at 325 MHz: 
24 February 2004 (solid line) and 21 December 2005 (dashed line). The top
panel shows the regular average profile, for comparison.}
\label{ch4_aft_null_ind}
\end{figure}
\end{center}
\begin{center}
\begin{figure}
\begin{center}
\includegraphics[angle=0,width=0.5\textwidth]{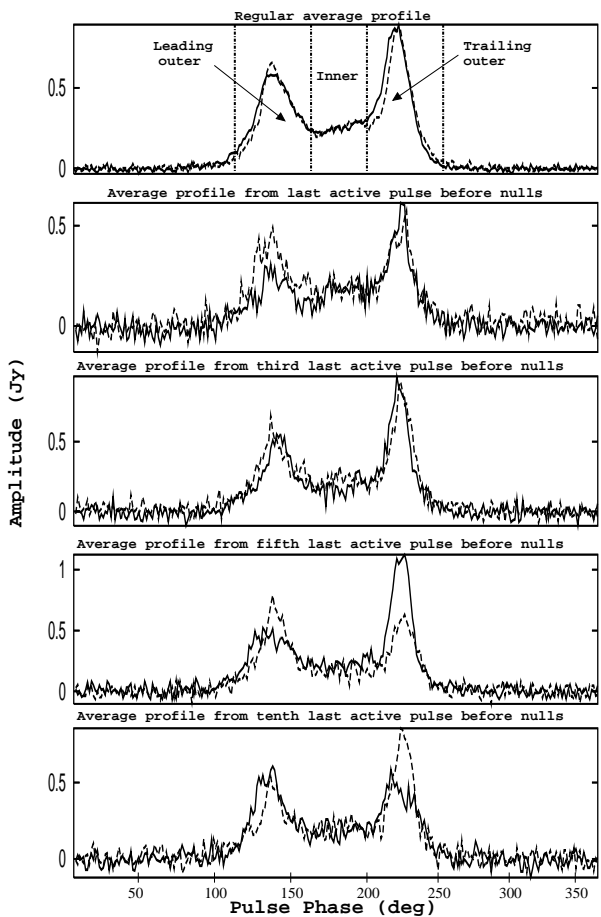}
\end{center}
\caption[Comparison of the average profile from first few actives pulses
immediately before the nulls at 325 MHz from observations on 24 February
2004 and on 21 December 2005]{Comparison of the average profile from the 
first, third, fifth and tenth from last pulse before the onset of nulls,
for the two epochs at 325 MHz: 24 February 2004 (solid line) and 21 December 
2005 (dashed line). The top panel shows the regular average profile, for comparison.}
\label{ch4_before_null_ind}
\end{figure}
\end{center}

We notice that these properties are pretty much the same for the two different 
epochs of 325 MHz data (e.g. Fig. \ref{ch4_1p_aft_null}), suggesting consistency
in the observed properties of this pulsar before and after the nulls. To investigate 
this further, we have compared the average profiles at 325 MHz from the two epochs, 
for successive pulses in the bursts and also for successive pulses before the onset 
of nulls. The results are shown in Figs. \ref{ch4_aft_null_ind} and Fig. \ref{ch4_before_null_ind}.
A noticeable similarity is observed between the corresponding average
profiles from the two epochs, for the pulses in the burst, whereas the
similarity is much less for the pulses before the onset of the nulls.
This further supports the claim that the pulsar shows very characteristic 
changes in profile intensity, shape and energy distribution when it
emerges from nulls. These changes are investigated further in detail
in the next section.

\begin{center}
\begin{figure*}
\begin{center}
\includegraphics[angle=0,width=1\textwidth]{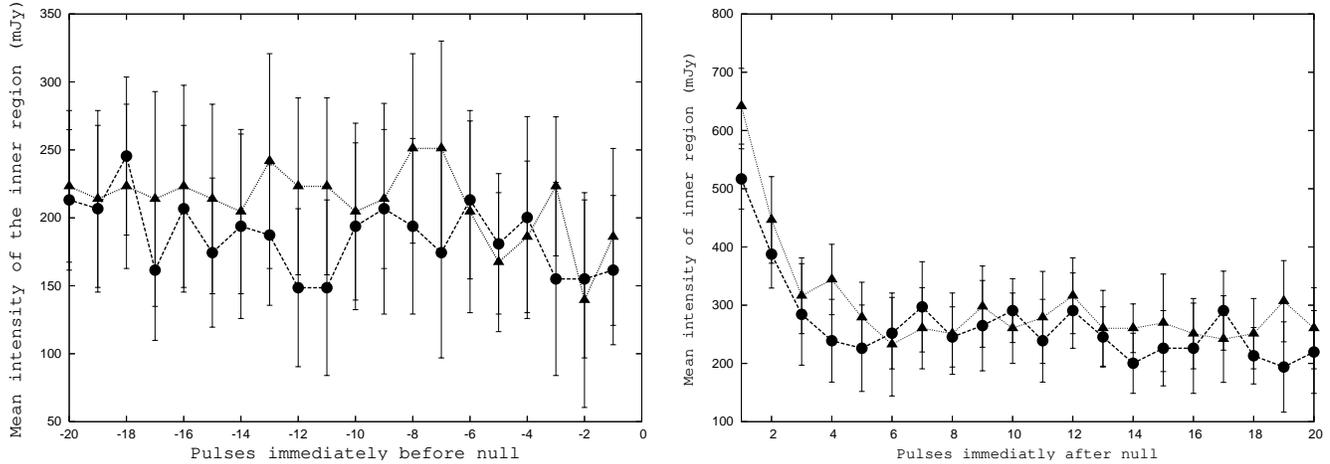}
\end{center}
\caption[Mean intensity of inner region of the average profile from the
pulses immediately before and after the nulls]{Left panel: Mean intensity of
inner region of the average profile from the pulses immediately before the
nulls versus the corresponding pulse number at 325 MHz. Right panel: Mean
intensity of inner region of the average profile from the pulses immediately
after the nulls versus the corresponding pulse number at 325 MHz. Results
for the epoch on 24 February 2004 are denoted by dashed line and for the
epoch on 21 December 2005 are denoted by dotted line.}
\label{ch4_mean_int_inner_bef_aft}
\end{figure*}
\end{center}


\subsubsection {Variation of average pulse intensity around the nulls}
\begin{center}
\begin{figure}
\begin{center}
\includegraphics[angle=0,width=0.5\textwidth]{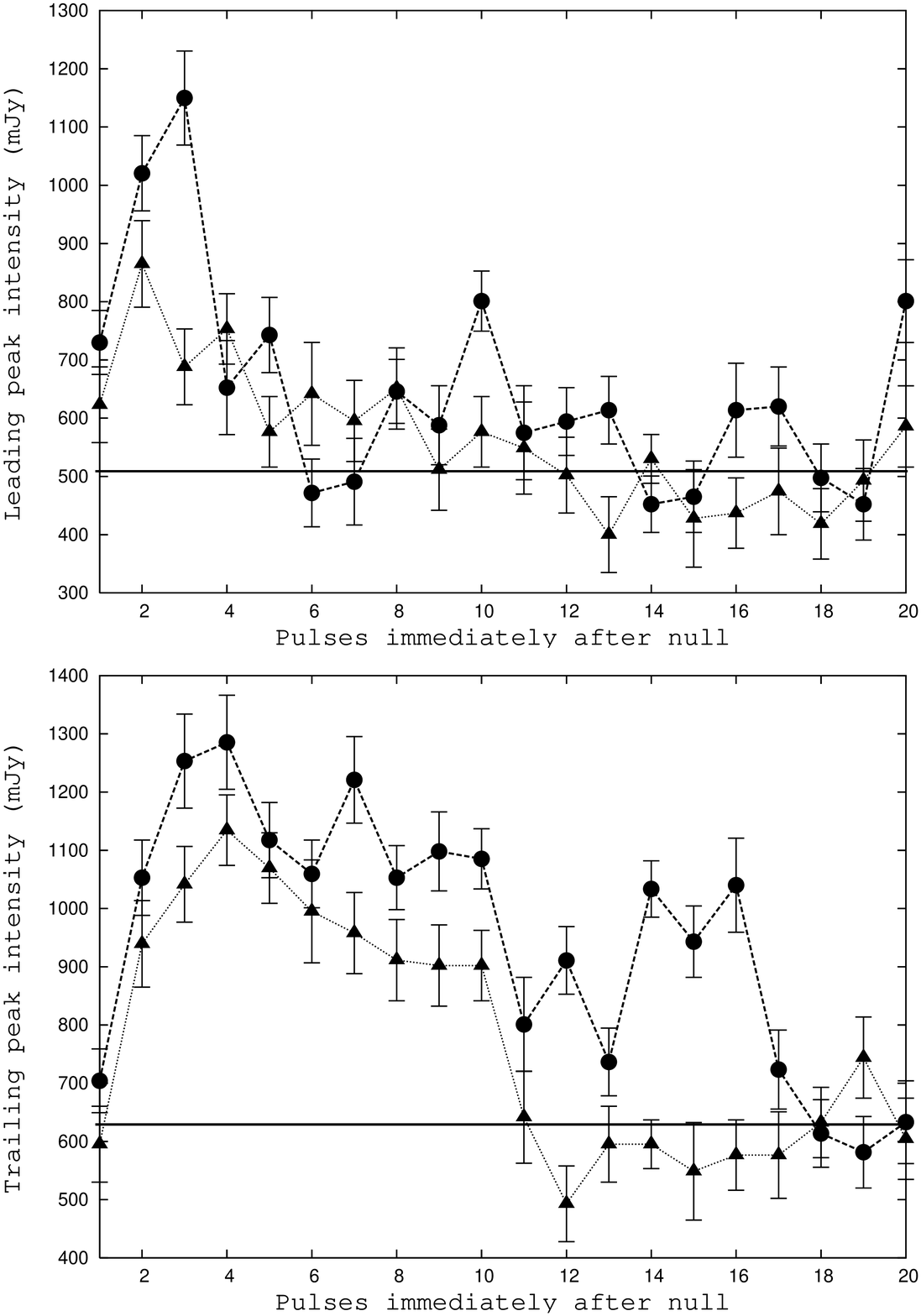}
\end{center}
\caption[Mean intensity of leading peak, trailing peak of the average
profiles from the pulses immediately after the nulls]{Top panel: Mean
intensity of the leading peak of the average profile from the pulses
immediately after the nulls versus the corresponding pulse number at 325
MHz. Bottom panel: same as the top panel, but for the trailing peak of the
average profile from the pulses immediately after the nulls. For both cases,
5 bins on each side of the peak have been included, covering a pulse 
longitude range of 13.5 degrees. In both panels, results from the data 
of 24 February 2004 are shown by dashed lines and that of 21 December 2005 
are shown by dotted lines; the horizontal solid lines indicate the corresponding 
values for the normal profile.}
\label{ch4_leading_trailing_ratio_int}
\end{figure}
\end{center}

In the following, we investigate in further detail the evolution of the
intensity in different regions of the pulse profile, of the pulses around the nulls.
Figs. \ref{ch4_mean_int_inner_bef_aft} and \ref{ch4_leading_trailing_ratio_int} show
the total intensity of the inner region and the leading and trailing outer
regions, for the average profiles obtained from the addition of specific
pulse numbers in the burst regions, as a function of the pulse number.
In addition, Fig. \ref{ch4_mean_int_inner_bef_aft} also shows the same
for pulses before the onset of the nulls, for the inner region only.  For
the inner region, during the bursts, the mean intensity follows a clear, 
systematic trend: it is maximum for the first active pulse in the bursts
($\sim$ 520 mJy for the first epoch and $\sim$ 650 mJy for the second epoch), 
and then gradually goes down.  
The intensity of the inner region reaches the value
observed for the normal profile ($\sim$ 220 mJy for the first epoch and 
$\sim$ 260 mJy for the second epoch) after about 20 active pulses in the 
burst\footnote{Peak to peak fluctuation of the mean off pulse intensity is used as the measure of
error. Corresponding error bars denote 3-sigma errors.}. It is notable that these variations of intensity with pulse number
is strikingly similar for the two epochs.

The corresponding behaviour for the pulses before the onset of the nulls is 
somewhat less clear. The mean intensity of the inner region is less for 
the average profiles from these pulses (around 164 mJy for the first epoch 
and 187 mJy for the second epoch), and also shows some signature of a gradual 
decrease from the 20th to the last pulse before the onset of nulls (left panel 
of Fig. \ref{ch4_mean_int_inner_bef_aft}) $-$ the value for the 20th pulse
(214 mJy and 224 mJy for the first and second epochs, respectively) is quite
close to the normal profile value.  However, this behaviour is not as clear
and systematic as that for the pulses in the bursts.  One possible reason for
this could be that as the decrease of intensity before the nulls happens over
slightly different timescales for each individual null, the averaging process
tends to blur out the trend.  That this does not happen for the pulses in the
bursts indicates that the behaviour of the pulsar as it emerges from the nulls
is probably highly repeatable.  It is thus likely that the pulsar magnetosphere
reaches a very similar condition during the nulls and the pulsar turns on in a
well defined, repeatable state after each null.


The variation of mean intensity for the leading and trailing outer regions
of the profile (Fig. \ref{ch4_leading_trailing_ratio_int}) is somewhat 
different from that of the inner region, for the pulses in the bursts.
Though the intensity at the beginning of the bursts is higher, the peak
is slightly delayed -- it occurs at the second or third active pulse in 
the bursts for the leading component, and at the third or fourth pulse for
the trailing component, rather than at the first pulse (as for the inner 
region). This delay can also be seen in Fig. \ref{ch4_pe_325_zoom} with 
careful observation. We would like to emphasize that the start and end of the 
nulls are defined by the inner region. The fall-off also appears to be 
somewhat more gradual than the 3-4 pulse decay seen for the inner region. The 
relative intensities of the leading and trailing outer regions also show a well 
defined behaviour. For the first pulse in the bursts, the two regions are of 
similar intensity (Fig. \ref{ch4_1p_aft_null}), and then the ratio evolves 
with pulse number and reaches the final value of 0.6 seen for the normal profile.

\begin{center}
\begin{figure}
\begin{center}
\includegraphics[angle=0, width=0.5\textwidth]{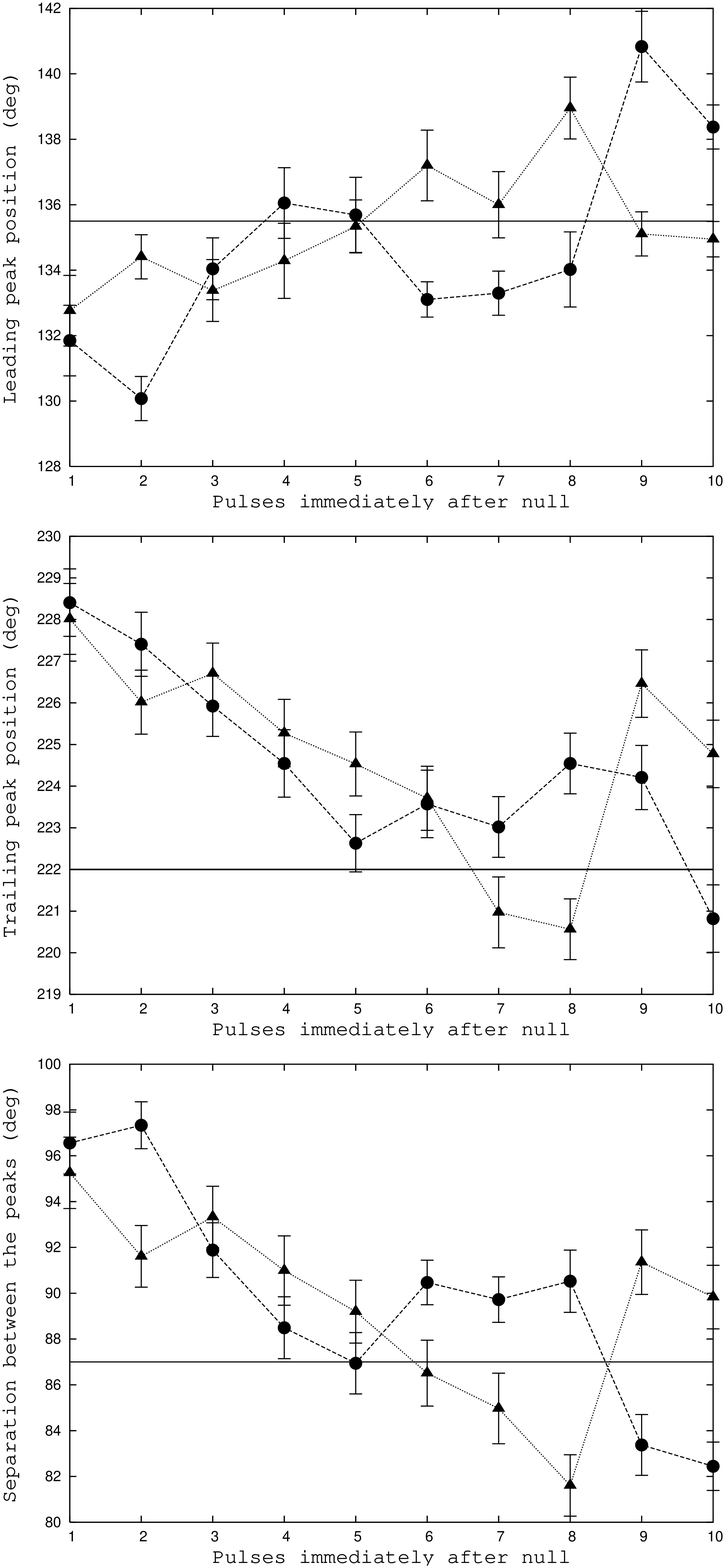}
\end{center}
\caption[Leading and trailing peak positions of the average profiles from
the pulses immediately after the nulls]{Top panel: Position of the leading
peak of the average profile from the pulses at the beginning of the burst,
versus the corresponding pulse number at 325 MHz. Middle panel: Position of
the trailing peak of the average profile from the pulses at the beginning of
the burst, versus the corresponding pulse number at 325 MHz. Bottom panel:
Separation between these leading and trailing peaks versus the corresponding
pulse number at 325 MHz.  In each panel, results from the data of 24 February
2004 are shown by dashed lines and that of 21 December 2005 are shown by dotted
lines; the horizontal solid lines indicate the corresponding values for the
normal profile.}
\label{ch4_l_t_pos_sep}
\end{figure}
\end{center}
\subsubsection {Variation of profile width of pulses around the nulls}

Fig. \ref{ch4_avp325_24feb_21dec} gives an indication that the positions of 
the leading and the trailing peaks of the average profile from the first 
pulses in the bursts are shifted from the normal.  To check if there is a 
systematic behaviour of this with pulse number in the bursts, we take two
consecutive active pulses in the bursts and calculate the mean average 
profile for these (e.g. average profiles from the first and second pulses,
third and fourth pulses etc.). The positions
of the leading and the trailing peaks are calculated for each of these mean
profiles by fitting second order polynomials to the peaks. We combine two
consecutive pulses, in order to increase the signal to noise of the resultant 
profile, which helps us fit a function to determine the position of the peak. 
Fig. \ref{ch4_l_t_pos_sep} shows the positions of the fitted peaks for the
leading and trailing sections as well as the separation between them, as a 
function of pulse number. The leading peak is shifted to earlier pulse phase
(by about 3 to 4 degrees) at the beginning of the burst and slowly comes back 
to the normal profile position in about 10 pulses.  The trailing peak is shifted
to later pulse phase by somewhat larger amounts (about 6 degrees) at the 
beginning of the burst, and it also comes back to the normal profile position
in about 10 pulses.  As a result, at the beginning of the burst, the profile 
width, as defined by the separation between these two peaks, is larger by about 
9 to 10 degrees from its normal value of 87 degrees, and relaxes to this normal
value in about 8 to 10 pulses.  Also, the center of the profile, defined by
the point mid-way between the two peaks, is displaced to later pulse phase at
the beginning of the burst (by about 2 degrees).   

The above findings, which are very similar for both the epochs of observations,
provide further evidence for a well defined change in the emission properties of 
the pulsar when it emerges from the nulls, and a systematic evolution of the 
same during the first few pulses of the bursts following the nulls.

\subsection{Behaviour of drift pattern around the nulls} \label{drift_around_null_1}

\begin{figure*}
\hbox{
 \hspace{1cm}
  \includegraphics[angle=0, width=0.42\textwidth]{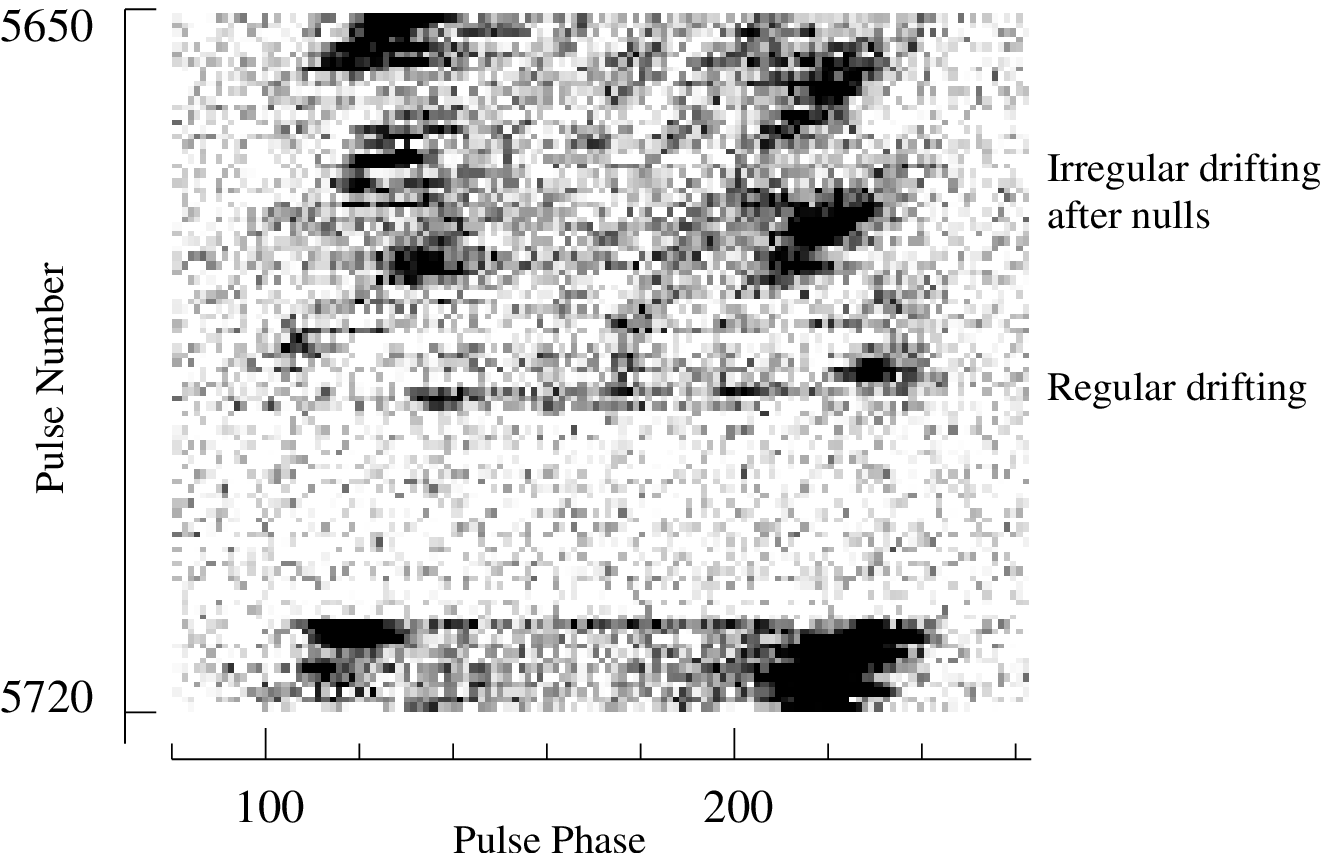}
  \includegraphics[angle=0, width=0.42\textwidth]{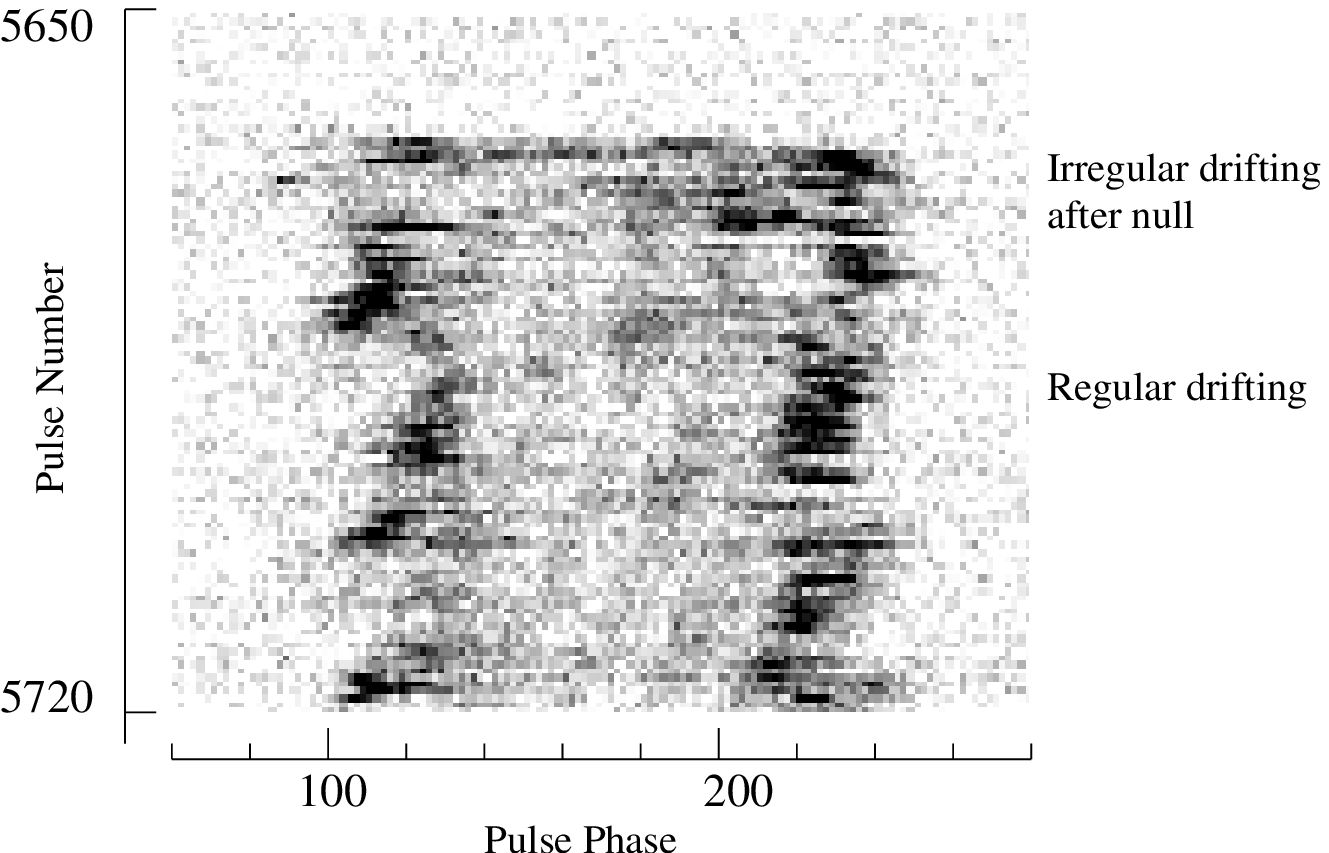}
   }
\caption{Gray scale plot of 70 single pulses of PSR B0818$-$41 at 325 MHz. 
Left panel: demonstrate the nature of drifting just before the nulls. 
Right panel: demonstrate the nature of drifting just after the nulls.}
\label{sp_before_after}
\end{figure*}

In \cite{Bhattacharyya_etal_09}, we have described the general behaviour of 
the drift pattern around the nulls for this pulsar.  There is clear evidence
for changes in the apparent drift rate just before and after the nulls.  On
several occasions, the apparent drift rate becomes less, i.e. drifting becomes
apparently slower, often transitioning to an almost phase stationary drift 
pattern just before the onset of the null (see left panel of 
Fig. \ref{sp_before_after} for a typical example). A detailed examination 
of the nulls selected in Sect. \ref{ch4_int_indnull} shows that this kind 
of behaviour is seen for $\sim$ 60\% of the occasions. For the remaining 
cases, there does not appear to be any appreciable change in the apparent drift rate. 

The behaviour of the drift pattern just when the pulsar comes out of the nulls
is also interesting.  It often shows irregular drifting for a few pulses and
then settles down to the normal drifting pattern (see right panel of 
Fig. \ref{sp_before_after} for a typical example).  This kind of 
transition is seen after most of the nulls. It is likely that there is 
some connection between the intensity changes that the pulsar undergoes 
just before and after the nulls and the changes in the drift pattern seen 
at those times.

\section{Discussion}\label{ch4_discussion} 
In the following we discuss about the implications of the new results from 
our study of the nulling properties of PSR B0818$-$41, and compare these 
with results reported in the literature for other pulsars.  We also 
investigate the relevance of the Partially Screened Gap model in 
understanding some of our results.

\subsection{Intensity modulation around the nulls} \label{intensity_around_nulls}

Our investigations reveal a clear difference in the nature of the transitions
from bursts to nulls to that from nulls to bursts.  The pulsar's radio energy
appears to reduce gradually for a few pulses before it finally switches off
at a null.  The time scale of this fading ranges from about 5 to 13 pulses, 
for different nulls.  On the other hand, the transitions from nulls to bursts
are abrupt and furthermore, the pulsar's radio energy for the first few pulses 
in a burst is significantly larger than the average value.  For the inner 
region, the brightening from just before the nulls to the beginning of the burst 
is, on an average, close to 3 times, at 325 MHz. This ratio appears to show 
some frequency dependence, dropping to close to 2 at 610 MHz.  For the outer
regions, the brightening is about 2 times at 325 MHz.  In addition, the 
brightening of the outer regions happens with a delay of 2-4 pulses from
the beginning of the burst, unlike the inner region where the first pulse
in the burst is the brightest and the intensity decays monotonically after
that.

It is interesting to check if other pulsars also show such a behaviour, or is
PSR B0818$-$41 unique in this context.  For PSR B0031$-$07, \cite{viv95} 
reported that both bursts and nulls are mostly quite abrupt, but in 
$\approx$ 20\% of cases the onset of nulling is slow, occurring over one to
several periods.  No significant relative brightening from before to after
nulls (in an average sense) is reported by them for this pulsar.  
Fig. 4 of \cite{rw08} plots pulse energy versus pulse
number for PSR J1819$+$1305, where we note a clear signature of gradual 
decrease in energy in the last active pulses before the onset of the nulls, 
and sharp transitions from nulls to bursts, with first pulses in the bursts
being much brighter than the normal pulses.  This is, however, not mentioned
and discussed by the authors in the paper.  PSR B1944$+$17 is another 
pulsar for which the transitions from nulls to bursts are known to be quite 
different in character from the transitions from bursts to nulls. \cite{dchr86} 
reported that the switching off of this pulsar is preceded by a slow 
($\sim$ 3 pulses) decay in average intensity.  They also find a higher 
relative brightness of the first active pulse in the bursts, compared to
the average strength.  For PSR B0809$+$74 also, \cite{la83} find that the 
last active pulse before nulls is dimmer than the normal pulses, and the first
active pulse in the bursts is brighter than normal.

Thus, it appears that there is some evidence in the general pulsar population for 
the kind of behaviour that we report for PSR B0818$-$41.  For some nulling pulsars, 
such systematic intensity modulations may be partially masked by modulations due to other
causes such as drifting subpulses, as we find for the outer regions of PSR B0818$-$41. 
For the inner region, where the grazing line of sight samples a larger portion of the 
emission region, subpulses from multiple drift bands are averaged and as a result, 
intensity modulations due to these are smoothed out. This warrants a more detailed 
study of the intensity modulations around nulls for other pulsars.

\subsection{Shape and width of pulses around the nulls} \label{shape_width_around_nulls}

We have reported significant evolution of the shapes of the pulses around
the nulls, especially at the beginning of the bursts.  Shape changes of 
the first few pulses in the bursts include the enhanced bump of intensity 
in the inner region, a more symmetric profile with the ratio of strengths 
of the leading and trailing components becoming close to unity.   This is 
accompanied by an increase of about 10\% in the width of the profile, as
well as a shift of the mid-point towards the trailing side.

Such effects are seen in some other pulsars also.  \cite{jv04} reported 
significantly different shapes of the average profiles from pulses before 
and after the nulls for PSR B0818$-$13 $-$ the average profile from the last 
and first active pulses immediately before and after the nulls are observed 
to be double peaked, whereas the normal average profile are observed to be 
single peaked.  For PSR B1944$+$17, \cite{dchr86} reported that the last 
pulse before null has a shape that is quantitatively different and more 
variable than the shapes of other pulses.  For PSR B0809$+$74, \citep{vsrr03}
found that though the pulses immediately before and after the nulls are 
similar in shape to the average profile, the peak of average profile from 
the first active pulse in the bursts after nulls is shifted towards earlier 
pulse longitude. As an explanation for this, \cite{vsrr03} proposed that 
after the nulls sub-beam carousel is smaller, indicating that we are looking 
deeper in the pulsar magnetosphere than we do normally.  Finally, as an 
exception, we mention the case of PSR B0031$-$07, for which 
\cite{viv95} did not find any significant difference in the average profiles
from the first active pulse in bursts, the last active pulse before the
nulls, and the normal average profile.

For PSR B0818$-$41, the change in pulse width and center of the profile that
is reported by us can be interpreted as a change in the distribution of the emission 
regions in the pulsar's magnetosphere. The pulse width can increase if 
either the cone of emission originates from a higher altitude on the same
set of field lines, or if it shifts to outer field lines while maintaining a
constant altitude.  A combination of both these effects is also possible.  
If a increase in height alone was responsible, then the mid-point should 
have moved to earlier longitudes, due to increased aberration and retardation
effects \citep{ganga_etal}. If a change of field line alone was responsible, 
then the mid-point should have remained at the same longitude. The observed 
behaviour appears to require a combination of a shift to more outer field lines 
along with a reduction in the emission height. The shift to an outer set of 
field lines needs to be such that it produces an increase in the pulse width 
which is somewhat larger than observed. Some of this increase would then be 
compensated by the reduced height of emission, which should also be enough to 
produce the observed shift of the mid-point to later longitudes.

\subsection{Drift rate around the nulls} \label{drift_around_nulls}

The above picture becomes even more interesting when we add the information
about the changing drift rates just before and after the nulls.  It appears
that as the pulsar's radio intensity dims gradually before the nulls, there
is often an accompanying reduction in the drift rate.  Further, when the
pulsar comes out of the nulls, the increased radio intensity for the first
few pulses is very often accompanied by what looks like a disturbed drift
rate behaviour, which settles down to the regular drift pattern as the pulsar
intensity returns to normal.  This correspondence of intensity behaviour with
drift rate variations, though not indisputably strong, is nevertheless quite
striking and suggests a common cause.  One such possibility is explored in
the next section.

What is known about this kind of property for other pulsars?
Investigating drifting around the null for PSR B0809$+$74, \cite{vsrr03}, found 
that drift rate just before a null does not deviate from the normal drift rate. 
However, drift rate just after nulls is different from the normal drift rate. After 
the nulls PSR B0809$+$74 goes to a quasi-stable mode. They also found that drift 
rates after longer nulls are lower than the normal average drift rate.

\bigskip
From the discussions in Sect. \ref{intensity_around_nulls}, \ref{shape_width_around_nulls} 
and \ref{drift_around_nulls}, it is clear that there are some very specific and 
well correlated changes that are seen in the radio emission properties of 
PSR B0818$-$41 when it restarts emission after a null. The fact that these 
changes are seen to be quite similar on two different epochs of observations
strongly supports that these are stable, intrinsic changes and are tightly 
coupled to the nulling process. This points strongly to a scenario where
the electromagnetic conditions in the region of the magnetosphere responsible
for the radio emission reach a well defined ``state'' during or towards the 
end of each null; in other words, some kind of a ``reset'' of the pulsar's 
radio emission engine takes place, as a result of which the pulsar starts with 
a very characteristic post-null behaviour that slowly evolves towards a 
different kind of behaviour which characterises the average properties of 
the pulsar. Just before the onset of null, though the behaviour is fairly
characteristic, it is not as stable or repeatable as that just after the nulls.
In the following we explain some of these results with Partially Screened Gap
model.

\subsection{Explanation with the Partially Screened Gap model} \label{sec:psg}
According to the widely accepted picture, each radio subpulse can be 
associated with a radio sub beam passing through our line of sight, which 
is emitted at some altitude (canonically about 50$-$100 stellar radii) 
within a plasma column that is directly related to a spark discharge within 
the inner acceleration region near the polar cap surface. 
The spark discharge produces a Goldreich Julian density of the primary (highly 
relativistic) column of plasma, which is then Sturrock multiplied and formed 
into a column of very dense but much less relativistic secondary plasma. 
Complicated non-linear processes in these plasma lead to the generation of 
the coherent radio emission associated with the observed subpulse 
(e.g. \cite{cheng_etal}, \cite{Filippenko_etal}, \cite{Melikidze_etal}, \cite{Mitra_etal}). 
There is no clear picture about how the energy density is propagating from the 
spark to the unstable secondary plasma and then to the radio emission beam. 
However, one can assume that the stronger the electric field in the acceleration 
region the stronger is the radio intensity of the observed subpulses. 
The prototype of the inner acceleration region was the pure vacuum gap model of 
\cite{Ruderman_etal}. It is now understood that this model predicts too fast a 
subpulse drift since the electric field in pure vacuum gap is too strong. Much 
better agreement with the observational data of drifting subpulses is achieved 
within the so-called Partially Screened Gap model (hereafter PSG) (\cite{Gil_etal_03}, 
\cite{Gil_etal_08}), in which the electric field is screened and lowered by factor 
of about 10 due to thermal emission of iron ions from the hot polar cap surface.  
The polar cap is heated due to the reverse flow of electrons from the magnetically 
created electron-positron pairs, even as the accelerated positrons leave the 
acceleration region.  The heating produces thermal ejection of ions from the 
surface, which partially screens the gap electric field. The PSG model postulates 
that the polar cap temperature, $T_s$, reaches a quasi-equilibrium value, slightly 
below the critical ion temperature, $T_i$. This quasi-equilibrium is established 
by a subtle thermostatic balance between the heating due to back-flow bombardment 
and cooling due to radiation: the higher the $T_{s}$, the larger the thermo-emission 
of positive ions, which screens the potential drop in the gap, reduces the acceleration
of the charges, thereby reducing the heating effect and thus leads to a decrease in 
the temperature (see \cite{Gil_etal_03} for details). In the equilibrium condition, 
the value of $T_{s}$ is about few percent lower than $T_{i}$ (which is about $10^6$ K), 
whereas the potential drop is only a few percent of the vacuum gap value (see 
Appendix in \cite{Gil_etal_08}). 

It is quite likely that the actual surface temperature $T_s$ varies slightly 
around a thermostatically determined value on relatively slow time scales of 
longer than pulsar period. These tiny variations with an amplitude of few thousands K 
can be crucial for the pulse nulling phenomenon. We speculate that a null occurs when the 
polar cap temperature raises by few thousands K to a value at which the potential drop is 
screened to a level making generation of detectable radio emission impossible. In this 
phase the remnant potential drop can still drive enough pair production to heat the polar 
cap surface but not enough to generate detectable radio emission higher up in the magnetosphere. 
This residual heating is very important and its actual amount will determine the time of 
null duration. Without a residual heating the nulls would be extremely short (below 100 ns;
\cite{Gil_etal_03}) and unnoticeable. After some time the radiation cooling prevails
and the temperature drops back few thousands K to the PSG thermostatic regime. The
pulsar is back in the normal emission mode. Most likely, before reaching the quasi-stationary
value the epoch of even lower temperatures will be reached, when the potential drop
is even higher and the drift is fast, aliased and chaotic.

The electric field in the inner acceleration region above the polar cap
results from deviations of the actual charge density from the co-rotational
Goldreich-Julian (1969) value. This "non-co-rotational field" has two natural
components: parallel and perpendicular to the surface magnetic field B. The
parallel component causes acceleration, and in consequence pair plasma
production which is utilized in the radio emission generation process
further away. Also the heating of the polar cap surface, including thermal
ions ejection, are due to this parallel component. Second component is
tangent to the surface of the polar cap. This component causes the spark
plasma circulation around the polar cap and in consequence subpulse drift
across the pulse window. It is obvious that any variation of the gap
electric field concerns both components. Therefore, the decrease/increase of
the subpulse intensity should be correlated with slowing down/speeding up of
the (non-aliased) subpulse drifting. In other words, the gradual decrease of
subpulse intensity observed just before a null should be associated with the
gradual slowdown of the subpulse drift. Exactly such a correlation is
reported in our paper and its explanation (at least qualitative) seems
clear. Just before a null the gap electric field decreases gradually on the
time scales of several to few tens of pulsar periods, which causes a gradual
drop of both the subpulse intensity and their drift rate. We speculate that
this gradual decrease of gap electric field is caused by a small increase of
the surface temperature which is estimated later in this section.

After the nulls the intensity rises to maximum over a short (less than one
period) time scale, keeping a high intensity for a number of pulses (often
randomly scattered over the pulse window). Then the normal drifting mode
begins, with intensity slightly lower than those just after the nulls. This
behaviour suggests strongly that at some point during a null the electric
field begins to rise again and as soon as it reaches a critical value the
PSG resumes an operation and the pulsar is back in action. Again, one can
speculate that this rise of the electric field is associated with a small
drop of the surface temperature. Right after the null the electric field is
probably stronger than in the stable normal drifting mode. This is why the
subpulses after a null are initially stronger than in the normal mode. We
believe that under this strong electric field the spark circulation is very
fast but also slowing down quickly as the electric field decreases. 
The system goes through fast variations of carousel circulation speed, which 
may result in the erratic behaviour of the observed subpulses.
Later, after few to several pulses the PSG reaches the stable state and resumes 
normal operation. The pulsar is back in the normal drifting mode.

Using the phase offset between the leading and the trailing outer regions, we
found that subpulse drifting in PSR B0818$-$41 is most likely first order
aliased \citep{Bhattacharyya_etal_09}. The carousel rotation period $P_4
\sim $ 18.3 $P_1 $ and time interval between recurrence of successive drift
bands at a given pulse longitude $P_3^t$=0.95 $P_1$. The observed slowing down of 
the drift rate to the point of phase stationary apparent drift bands just before 
the nulls (discussed in Sect. \ref{drift_around_null_1}) means that $P_3^t$ is 
close to $P_1$. Hence, just before the nulls $P_3^t$ must increase by about 5\% 
(from 0.95 to about 1.0). This small variation of drift rate could be due to small 
change in the neutron star's surface temperature \citep{Gil_etal_03}. In the 
following we estimate required change in polar cap temperature to increase 
$P_3^t$ by 5\%. Within the PSG model the actual potential drop is set at the level 
of few percent of the maximum possible \cite{Ruderman_etal} value, i.e. 
$\Delta V=\eta V_{RS}$, where the screening factor 
$\eta=1-\rho_{i}/\rho_{GJ}=1-exp[C(1-T_{i}/T_{s})]$ and the coefficient $C$ depends 
on the chemical composition of the surface layer (\cite{ml07}, \cite{Gil_etal_08}). 
In the carousel scenario of the subpulse drift based on the PSG model $P_{4}\sim \eta^{-1}r_{p}/h=NP_3^t$, 
where $r_{p}$ is the polar cap radius, $h$ is the height of the acceleration 
region and $N\sim 2\pi r_{p}/(2h)=\pi r_{p}/h$ is the number of sparks circulating 
around the polar cap. Thus, the basic drifting periodicity $P_3^t=\eta ^{-1}/\pi =0.318/\eta $. 
Assuming that ion critical temperature $T_{i}=2\times 10^{6}$ K (which seems to be a 
typical value; \cite{Gil_etal_08}), for change of $P_3^t$ from 0.95 to 1.0; the 
screening factor $\eta$ must change from 0.335 to 0.318 and hence the corresponding 
ratios $T_{s}/T_{i}$, will be equal to 0.98658 and 0.9874, respectively. The required 
change in the surface temperature is $\Delta T_{s}=0.00082T_{i}=1640$ K (from 1.973 to 
1.9748 MK). So, to explain 5\% slow-down of drift in B0818$-$41 just before the onset 
of the nulls, one would have to invoke about 0.1 \% polar cap surface temperature 
variation. Based on the PSG model \cite{Gupta_etal} argued (see their section 4.4) that 
only about 0.14 \% change in the surface polar cap temperature (i.e. a change of about 
4000 K around 2$\times 10^6$ K average temperature) is needed to cause this 8 \% slow-down 
of drift signatures in B0826$-$34; which is of the same ballpark to our estimation for B0818$-$41.

Based on the frequency evolution of average profile, the observed PA swing and the
analysis of subpulse drift signatures we proposed two possible geometries for
this pulsar, \textbf{G-1} ($\alpha=11\degr$, $\beta =-5.4\degr$), and \textbf{G-2}
($\alpha =174.5\degr$, $\beta =-6.9\degr$) in \cite{Bhattacharyya_etal_09}. Pulsar 
radiation pattern simulated with both the geometries reproduces the observed 
features, except for some differences. Although \textbf{G-2} provides a reasonable 
fit to the overall PA curve at 325 MHz, and the left half of the PA curve at 610 MHz,
only the middle part of the PA curve at 610 MHz can be fitted with \textbf{G-1}. On 
the other hand \textbf{G-1} appears to give better match on the observed $P_{2}^{m}$ 
(longitude separation between the drift bands) values and the over all drift pattern. 
Both the geometries can explain the slight changes of apparent drift rates and phase 
stationary drift bands, once aliasing of drifting subpulses is considered. The observed 
slow down of apparent drift rate before nulling can be interpreted as slowing down of 
the carousel in \textbf{G-1} and speeding up of the carousel in \textbf{G-2} (see 
Sect. 7.1 of \cite{Bhattacharyya_etal_09}). However, within the PSG model the 
accelerating gap electric field must increase to cause the speed up of the sparking 
carousel; and electric field should decrease to cause the slow-down of the same. A 
correlation between the gradual slowdown of the drift rate and gradual decrease of 
pulse energy is expected within the inner gap acceleration model. Hence, the 
fact that we do observe the gradual intensity decrease over a number of pulses just 
before the nulls strongly suggests that the accelerating electric field decreases 
in this stage of pulsar activity. This automatically implies that the carousel 
slows-down before the nulls, which is the case for \textbf{G-1}. Thus comparison 
of the prediction of PSG model with observations favors \textbf{G-1} over \textbf{G-2}.

\section{Summary} \label{sec:summary}

Following are the interesting new results from our investigation of the nulling 
properties of PSR B0818$-$41.\newline
{$\bullet$ The pulsar shows well defined nulls, lasting in duration from a few tens
of pulses to a few hundreds of pulses; the estimated nulling fraction at 325 MHz
is about 30\%.}\newline
{$\bullet$ The transitions from bursts to nulls are gradual ($\sim$ 10 pulse 
period on average), whereas the transitions from nulls to bursts are rather abrupt 
(less than one pulse period). The last few active pulses before the nulls are less 
intense than the normal, whereas the first few active pulses just after the nulls 
outshine the normal pulses.  This effect is more evident for the inner region $-$ 
the first few active pulses just after the nulls are about 2.8 times more intense 
than those before the nulls. Before the nulls, the intensity decreases gradually over 
about 10 pulsar periods, during which generally the apparent drift rate slows down 
(often showing apparent longitude stationary drift bands just before the nulls). 
At the beginning of the bursts, the intensity rises to maximum 
in less than one period, keeping a high intensity for a number of pulses (often 
randomly scattered over the pulse window). Then the normal drifting mode begins, 
with intensity slightly lower than that just at the beginning of the burst.}\newline
{$\bullet$ We observe significant evolution of the shapes of the pulses at the 
beginning of the bursts.  Average profile from the first active pulses in the bursts 
has a significant bump of enhanced energy for the inner region, and comparable strengths 
for the leading and trailing outer regions, quite unlike the normal profile. The width 
of the average profile from the pulses just after the nulls is about 10\% more than 
that of the normal.  This is accompanied by a shift of the profile mid-point towards 
the trailing side.  Some of these effects can be explained by a shift of the
emission regions to different heights and/or somewhat outer field lines in the 
pulsar magnetosphere.} \newline
{$\bullet$ We observe a very characteristic post-null behaviour of PSR B0818$-$41 when it 
restarts emission after a null. This indicates that some kind of ``reset'' of the pulsar's 
radio emission engine takes place during the nulls, and immediately after the nulls 
the conditions of magnetosphere responsible for the radio emission are well defined.}
\newline

The results presented in this paper indicates that phenomenon of nulling is intrinsic to 
the pulsar radio emission that is systematically ceased during the nulls. 
This study will put constraint on the models explaining pulsar radio emission and nulling. 
We successfully explained many of our results with the help of PSG model. However, some 
remains to be explained, for example, why nulling is associated with fast rise and slow fall 
of intensity, why intensity distribution across the profile immediately after the nulls is 
different than typical. More importantly, how the gap electric field varies around the nulls 
and what could actually cause a radio null, remains to be explained. 
Considering the PSG model, the actual mechanism of generation of the coherent radio emission 
must be very sensitive to tiny changes of the input parameters. 
We intend to devote a separate paper to this and the other problems mentioned above.
Clearly, investigations of the pre and post null emission properties of PSR B0818$-$41, presented
in this paper, emphasize that nulling provides an useful tool to probe the pulsar radio emission.
Though, there is some evidence in general pulsar population for the kind of behaviour that we 
report for PSR B0818$-$41, for some pulsars such systematic modulations may be partially masked 
by modulations due to other causes such as drifting subpulses. This warrants a more detailed 
study of emission properties around the nulls for other pulsars.

\bigskip
\noindent{\large\bf Acknowledgments:}
We thank the staff of the GMRT for help with the observations. The GMRT is 
run by the National Centre for Radio Astrophysics of the Tata Institute of 
Fundamental Research. BB thanks Ramesh Bhat for his comments at very early 
stage of the work. We would like to thank our referee Patrick Weltevrede 
for his suggestions which had improved the paper. JG acknowledges a partial 
support of the Polish Research Grant N N203 2738 33.



\label{lastpage}

\end{document}